\documentclass[pdflatex,sn-mathphys-num]{sn-jnl}%

\usepackage{graphicx}%
\usepackage{multirow}%
\usepackage{amsmath,amssymb,amsfonts}%
\usepackage{amsthm}%
\numberwithin{equation}{section}
\usepackage{mathrsfs}%
\usepackage[title]{appendix}%
\usepackage{xcolor}%
\usepackage{textcomp}%
\usepackage{manyfoot}%
\usepackage{booktabs}%
\usepackage{algorithm}%
\usepackage{algorithmicx}%
\usepackage{algpseudocode}%
\usepackage{listings}%
\usepackage[english]{babel}
\usepackage{csquotes}

\theoremstyle{thmstyleone}%
\newtheorem{theorem}{Theorem}
\newtheorem{proposition}[theorem]{Proposition}%
\newtheorem{corollary}[theorem]{Corollary}%
\newtheorem{lemma}[theorem]{Lemma}%
\theoremstyle{thmstyletwo}%
\newtheorem{remark}{Remark}%
\theoremstyle{thmstylethree}%
\newtheorem{definition}{Definition}%

\usepackage{babel}
\usepackage{float}
\usepackage{mathtools}
\usepackage{bm}
\usepackage{amsmath}
\usepackage{amsthm}
\usepackage{amssymb}
\usepackage{graphicx}

\usepackage{amsthm}
\usepackage{amsfonts}
\usepackage{comment}

\usepackage{subfiles}
\usepackage{pgfplots}
\pgfplotsset{/pgf/number format/use comma,compat=newest}
\usepackage{url}
\usepackage{tikz}
\usepackage{notations}
\usepackage{bbm}
\usepackage{bm}
\usepackage{floatflt}
\usepackage{subfiles}

\usetikzlibrary{arrows,decorations.pathmorphing,backgrounds,positioning,fit,petri}

\usepgfplotslibrary{fillbetween}

\newcommand{\E}{\mathbb{E}}
\newcommand{\R}{\mathbb{R}}
\newcommand{\N}{\mathbb{N}}

\renewcommand{\d}{\mathrm{d}}

\abstract{
We study multiscale  measures, defined by a hierarchy of partial Gibbs equilibria, from variational and probabilistic perspectives. We first show that they are the unique solutions of a generalized maximum-entropy principle in which conditional entropies at different levels carry different weights. We then construct a hierarchical reinforced multinomial process proving  a large-deviation principle for its empirical measure. The stochastic process leaves the typical empirical distribution unchanged but modifies the exponential cost of its fluctuations. The resulting rate function is a weighted sum of conditional Kullback–Leibler divergences, establishing a direct link with the entropic principle above. The relation between the reinforcement  and  Poisson-Dirichlet processes is discussed.}

\begin{document}
\title{From entropic constraints to reinforced processes: a probabilistic origin of multiscale measures}

\author[1]{Francesco Camilli}
\author*[1]{Pierluigi Contucci}\email{ pierluigi.contucci@unibo.it}
\author[1]{Emanuele Mingione}

\affil[1]{Department of Mathematics, University of Bologna, Piazza di Porta S. Donato, Bologna, Italy}



\maketitle


\section{Introduction}

The focus of this work is on a class of probability measures, referred to as
\emph{multiscale measures}. These extend the Boltzmann--Gibbs equilibrium
by incorporating a hierarchy of partial equilibria occurring at different temperatures,
or equivalently at different energy scales.
They arise naturally in systems characterized by a strong separation of time scales
and are widely used in several contexts \cite{Pre13}, in particular: renormalization group \cite{Polchi84,Gallavotti}, 
spin glass theory
\cite{MPV, leuzzi2007thermodynamics, Guerra2002BrokenRS, Tala_vol1,panchenko2015sherrington,MultiscaleSK_CM2018,TalASS,Barra11,camilli2025phase},
out of equilibrium dynamics \cite{Allahverdyan_2000pre, Kurchan-Cugliandolo,PMPF, Contucci_2019,Contucci_2021,
Vulpiani17,Cugliandolo_2011,Coolen_1993,alberici2024convergence,Ram22}, averaging theory \cite{Pavliotis}, inference and machine learning \cite{survey_transfer,TransferPRLFede,Gerace_2022}. The emergence of separated scales is intuitive in the first three contexts, where it reflects a genuine physical or dynamical hierarchy of relaxation times; in inference and machine learning the separation is instead typically \emph{informational} or \emph{architectural} rather than temporal. For instance, in transfer learning \cite{survey_transfer,TransferPRLFede,Gerace_2022} a model is first fit on a source task or a shared representation (the coarse, slowly-varying level) and only subsequently fine-tuned or adapted to a target task (the fast, fine-grained level), with the two stages naturally carrying different effective amounts of data and hence different effective ``temperatures'' in the sense of \S\ref{sec:twotemperatures}; similarly, in hierarchical or composite Bayesian models \cite{Pre13,Ram22} a population-level (coarse) parameter shared across many groups is estimated from pooled data, while group-specific (fine) parameters are estimated from far smaller, group-restricted samples, producing exactly the kind of asymmetric statistical resolution across levels that characterize a multiscale measure.{ Instances of multiscale measure in spin glass theory and machine learning are discussed in more detail in \S\ref{sec:related-models}.  In this work, we highlight the statistical-mechanical and information-theoretic structure of these objects, with particular emphasis on their probabilistic origins. 

We begin by recalling the standard construction of a multiscale measure.
Consider a system whose configuration space is a Cartesian product
\begin{align*}
X_r \times \cdots \times X_1 = X \ni \bx \mapsto H(\bx) \in \mathbb{R},
\end{align*}to which we assign a cost function or Hamiltonian $H$, and fix a collection of positive parameters
$\boldsymbol{\zeta} = (\zeta_r,\ldots,\zeta_1)$.
A multiscale measure is obtained through a sequence of nested Boltzmann--Gibbs
averages, each associated with an energy scale $\zeta_\ell$.
For $\ell = 1,\ldots,r$, the set $X_\ell$ indexes degrees of freedom evolving on a
characteristic time scale. The variables in $X_r$ are assumed to equilibrate first, at scale $\zeta_r$, while the remaining degrees of freedom are frozen. Accordingly, one postulates a Boltzmann--Gibbs distribution on $X_r$
proportional to $e^{\zeta_r H}$, with normalization
$\mathcal{Z}_{r-1}=\sum_{x_r} e^{\zeta_r H}$.
The variables in $X_{r-1}$ then equilibrate at scale $\zeta_{r-1}$ under the
effective potential $\log \mathcal{Z}_{r-1}$.
Iterating this procedure yields a cascade of partial equilibria,
\be
H(\bx)\,\, \xrightarrow{X_r}\,\,\frac{e^{\zeta_r H(\bx)}}{\mathcal Z_{r-1}}
\,\,\xrightarrow{X_{r-1}}\,\,
\frac{e^{\frac{\zeta_{r-1}}{\zeta_r}\log \mathcal Z_{r-1}}}{\mathcal Z_{r-2}}
\,\,\xrightarrow{X_{r-2}}\ldots
\ee
consistently with the assumed separation of time scales.

A complementary and particularly transparent interpretation is obtained by exploiting
the natural hierarchical structure of the decomposition
$X= X_r\times\cdots\times X_1$.
Each configuration corresponds to a leaf of a rooted tree of depth $r$,
while the intermediate nodes $(x_\ell,\ldots,x_1)$ with $\ell<r$ represent progressively coarser clusters obtained by grouping configurations level by level (see Fig.~\ref{fig:cluster-tree}).
Within this representation, the Hamiltonian $H$ encodes interactions between clusters and their descendants, and the multiscale measure assigns to each leaf a weight accumulated along its path to the root, with contributions coming from each level of the hierarchy.

\begin{figure}
    \centering
    \includegraphics[width=0.75\linewidth]{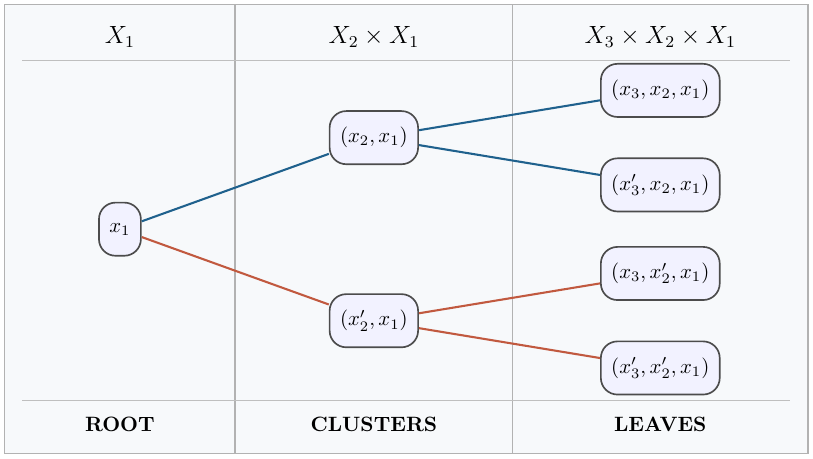}
    \caption{Hierarchical tree of depth $r=3$.
    An element (leaf of the tree) $\bx=(x_3,x_2,x_1)\in X$ uniquely determines a path
    reaching the root cluster; intermediate nodes represent clusters.}
    \label{fig:cluster-tree}
\end{figure}

The aim of this work is to  bring together variational and probabilistic viewpoints on this class of  measures. 
On the variational side, we show that the multiscale measure is the unique maximizer of a maximum-entropy functional in which the conditional entropies of the different levels carry different weights; in the two-level case the corresponding Lagrange multiplier is the ratio of the two inverse temperatures, and each level satisfies a fluctuation--dissipation relation at its own temperature. On the probabilistic side, our main contribution is the introduction of a hierarchical reinforced multinomial process whose large-deviation principle reproduces the entropic structure underlying multiscale equilibrium. 

In contrast with classical P\'olya-type schemes, where reinforcement typically changes the limiting empirical distribution \cite{Pemantle2007,POLYA,Muliere2000,FortiniPetrone2012}, the reinforcement considered here leaves the law of large numbers unchanged: the empirical measure still converges to the prescribed base measure $q$, and the associated rate function retains its minimum at $p=q$. What changes is the exponential cost of fluctuations away from this typical behavior. By modifying this cost separately at each level of the hierarchy and introducing the corresponding energetic tilt, the resulting variational problem selects precisely the multiscale Gibbs measure. The rescaling of the population between levels is deterministic, and this is not a matter of convenience: we show that a random rescaling, although it preserves the same law of large numbers, produces a rate function that is not of the required form. Related law-of-large-numbers-neutral reinforcement mechanisms have appeared in \cite{BudhirajaWaterbury2022,BertoinMallein2024I,BertoinMallein2025II,BertoinMallein2025III}; our construction uses this principle across nested scales in order to recover the multiscale structure, with an explicit rate function.

The paper is organized as follows. Section~\ref{sec:defintion} introduces multiscale measures, their generating functions and linear response identities, and recalls their role in statistical mechanics and learning. Section~\ref{sec:varprinc} characterizes these measures through a maximum-entropy principle with constraints on conditional entropies, and discusses their emergence from two-temperature thermodynamics. Section~\ref{sec:large-deviation-perspective} introduces the hierarchical reinforced multinomial process and proves the corresponding large-deviation principle; it also explains why random, rather than deterministic, population rescaling does not generally reproduce the required multiscale structure. Finally, Section~\ref{sec:chinese} discusses the Poisson--Dirichlet representation of two-scale measures and its realization through a Chinese restaurant process followed by ordinary multinomial sampling.

\section{Definitions}\label{sec:defintion}
We fix an integer $r\geq 1$ and a collection $\{X_\ell,\ell\leq r\}$ of finite sets, for the sake of simplicity. We denote by $X=X_r\times\cdots \times X_1$ the cartesian product of the collection, and its elements by $ \bx=(x_{\ell})_{\ell\leq r}\in X$. Given a  probability distribution $p$ on $X$,  for $1\leq \ell\leq r$, we set

\be\label{defconditionalpro}
p^{<\ell}\equiv p^{<\ell}(x_{\ell}|x_{\ell-1}\ldots,x_1):=\mathbb{P}(x_{\ell}|x_{\ell-1},\ldots,x_1)\,,\quad p^{<1}=\mathbb{P}(x_1)\,.
\ee 
We will refer to $p^{<\ell}$ as the conditional probabilities of the $\ell$-th level. The chain rule of conditional probabilities implies
\be\label{eq:probdecomposition}
p(\bx)=\prod_{\ell=1}^r p^{<\ell}(x_{\ell}|x_{\ell-1}\ldots,x_1)\,,\quad \bx\in X\,.
\ee
Let us also introduce
\be\label{def:marginal}
p^{(\ell)}(x_{\ell},\ldots,x_1):=\sum_{x_r,\ldots, x_{\ell+1}} p(\bx)\,,\quad \ell\leq r
\ee
namely the marginal probability distribution of the first $\ell$ variables. Hence  
\be
p^{(\ell)}=\prod_{k=1}^{\ell}p^{<k}
\ee
In the following $\langle\,\rangle , \langle\,\rangle_{<\ell},\langle\,\rangle_\ell$ denote the averages w.r.t. $p,p^{<\ell}$ and $p^{(\ell)}$ respectively. We are now in place to define a multiscale measure.

\begin{definition}[Multiscale measure]\label{def:multiscale measure}
Let $\bzeta:=(\zeta_{\ell})_{\ell\leq r}$ be a collection of positive real numbers, and $H:X\to\mathbb{R}$ a function called Hamiltonian or cost. Define the following backward recursion: 
\be\label{eq:recupartion}
Z_r(\bx) :=e^{H(\bx)}\text{ and } Z^{\zeta_{\ell}}_{\ell-1}(x_{\ell-1},\dots,x_1)=\sum_{x_\ell} Z_{\ell}^{\zeta_{\ell}}(x_{\ell},\dots,x_1),\quad \quad 1\leq\ell\leq r\,.
\ee
The multiscale measure associated with $(X,\bzeta,H)$ is then defined by the set of conditional probabilities
\be\label{eq:condprob}
p^{<\ell}\,=\,\dfrac {e^{\zeta_{\ell} \log Z_{\ell}}}{e^{\zeta_{\ell} \log Z_{\ell-1}}}\,\quad 1\leq\ell\leq r
\ee through the chain rule \eqref{eq:probdecomposition}. The normalization of $p^{<\ell}$ will be denoted by calligraphic $\mathcal{Z}_{\ell-1}:= Z_{\ell-1}^{\zeta_{\ell}}$. 

\end{definition}

\subsection{Generating functions}\label{sec:genfun}
In this section we discuss the statistical mechanics deriving from Definition \ref{def:multiscale measure}.

\begin{definition}[Multiscale pressure]
The $\ell$-th level generating function, or pressure, is
\be\label{eq:genfunc}
P_{\ell}= \log Z_{\ell},\quad  0\leq \ell\leq r
\ee
or equivalently, by the recursion \eqref{eq:recupartion},
\be\label{eq:rec press}
e^{\zeta_\ell P_{\ell-1}}=\sum_{x_\ell} e^{\zeta_\ell P_{\ell}},\,\quad 1\leq\ell\leq r \,.
\ee
\end{definition}

\begin{remark}\label{rem:physconv}

We notice that by setting $\zeta_r=1$ and $H\to-\beta H$, for some $\beta>0$
representing the inverse temperature of a physical system, the multiscale measure
can be seen as a cascade of $r$ Boltzmann-Gibbs averages. In order to highlight
the physical content it is convenient to set $\zeta_{\ell}=\frac{\beta_{\ell}}{\beta}$
for any $1\leq \ell\leq r$. The parameter $\beta_{\ell}$ represents the inverse
temperature of the $\ell$-th level of degrees of freedom, and the associated
$\ell$-th level free energy is $F_{\ell}=-\frac{1}{\beta} P_{\ell}\,$. The conditional probabilities \eqref{eq:condprob} can then be rewritten as
\be\label{eq:condprob2}
p^{<\ell}= e^{-\beta_\ell(F_\ell-F_{\ell-1})}\,\quad 1\leq \ell\leq  r\,.
\ee
From relation \eqref{eq:condprob2} one can interpret $p^{<\ell}$ as the canonical
Boltzmann-Gibbs measure on the set $X_\ell$ associated to an effective
Hamiltonian $F_{\ell}$ and inverse temperature $\beta_{\ell}$ where the degrees
of freedom $(x_{\ell'})_{\ell'>\ell}$ have been averaged out while the remaining
$(x_{\ell'})_{\ell'<\ell}$ are frozen.
\end{remark}

We will show that $P_{\ell}$ is the generating function of the $\ell$-th level conditional probability $p^{<\ell}$. We think of $P_\ell$ as a functional of the Hamiltonian, $P_\ell:\R^{X}\to\R$,
$H\mapsto P_\ell(H)$, the scale parameters $\bzeta$ being fixed. 
Let  $f:X\to\R$ and define
\be\label{eq:exprecursion}
f_r=f,\qquad
f_{\ell-1}=\langle f_\ell\rangle_{<\ell},
\qquad \ell=r,\ldots,1.
\ee
The conditional covariance at level $\ell$ is defined as
\be\label{eq:condcov}
\mathrm{Cov}_{<\ell}(f,g)
  :=\big\langle f_\ell\,g_\ell\big\rangle_{<\ell}-f_{\ell-1}\,g_{\ell-1}\,,
\ee

\begin{proposition}
\label{lem:pressure}
For all $f,g:X\to\R$ one has 
\begin{enumerate}
\item[(i)] the first variation of the pressures is the conditional expectation,
\be\label{eq:firstvar}
\frac{\d}{\d\lambda}P_\ell(H+\lambda f)\Big|_{\lambda=0}= f_{\ell}\,,\qquad 0\leq\ell\leq r\,;
\ee
\item[(ii)] the gradient of $P_0$ is the multiscale measure,
\be\label{eq:gradient}
\frac{\partial P_0}{\partial H(\bx)}=p(\bx)\,,\qquad \bx\in X\,;
\ee
\item[(iii)] the Hessian of $P_0$ is the hierarchical covariance form, each level
weighted by its own scale parameter,
\be\label{eq:hessian}
\sum_{\bx,\by\in X}\frac{\partial^2P_0}{\partial H(\bx)\,\partial H(\by)}f(\bx)g(\by)
=\sum_{\ell=1}^{r}\zeta_\ell\,\Big\langle\,\mathrm{Cov}_{<\ell}(f,g)\,\Big\rangle_{\ell-1}
\ee
\end{enumerate}
\end{proposition}

\begin{proof}
Write $P_\ell^\lambda$ for the pressures associated with $H+\lambda f$. Since $P_r^\lambda=H+\lambda f$, one has
$\frac{\d}{\d\lambda} P_r^\lambda\Big|_{\lambda=0}=f=f_r$. Differentiating the pressure recursion gives
\begin{align}
\frac{\d}{\d\lambda}P_{\ell-1}^\lambda\Big|_{\lambda=0}
=
\sum_{x_\ell}p^{<\ell}
\frac{\d}{\d\lambda}P_\ell^\lambda\Big|_{\lambda=0}= \left\langle\frac{\d}{\d\lambda}P_\ell^\lambda\Big|_{\lambda=0}\right\rangle_{<\ell}.
\end{align}
Therefore, a backward induction impies that for  any $0\leq\ell\leq r$ one has 
$\frac{\d}{\d\lambda}P_\ell^\lambda\Big|_{\lambda=0}=f_\ell$ which proves (i). In particular, $\frac{\d}{\d\lambda}P_0(H+\lambda f)\Big|_{\lambda=0}
=\langle f\rangle $ and taking $f=\delta_{\bx}$ one gets (ii). For (iii), differentiating the pressure recursion twice with respect to
$\lambda$ and using (i)  gives
\begin{align}
\frac{\d^2 P_{\ell-1}^\lambda}{\d\lambda^2}\Big|_{\lambda=0}
&=
\zeta_\ell
\left[
\left\langle
\left(
\frac{\d P_\ell^\lambda}{\d\lambda}\Big|_{\lambda=0}
\right)^2
\right\rangle_{<\ell}
-
\left(
\left\langle\frac{\d P_{\ell}^\lambda}{\d\lambda}\Big|_{\lambda=0}
\right\rangle_{<\ell}\right)^2
\right]
+
\left\langle
\frac{\d^2 P_\ell^\lambda}{\d\lambda^2}\Big|_{\lambda=0}
\right\rangle_{<\ell}\\
&=
\zeta_\ell
\left(
\big\langle f_{\ell}^2\big\rangle_{<\ell}
-\left\langle f_\ell\right\rangle_{<\ell}^2
\right)
+
\left\langle
\frac{\d^2 P_\ell^\lambda}{\d\lambda^2}\Big|_{\lambda=0}
\right\rangle_{<\ell}
=
\zeta_\ell\,\mathrm{Var}_{<\ell}(f)
+
\left\langle
\frac{\d^2 P_\ell^\lambda}{\d\lambda^2}\Big|_{\lambda=0}
\right\rangle_{<\ell},
\end{align}
where
\begin{align}
\mathrm{Var}_{<\ell}(f)
&:=
\left(
\big\langle f_{\ell}^2\big\rangle_{<\ell}
-\left\langle f_\ell\right\rangle_{<\ell}^2
\right).
\end{align}
Since $P_r^\lambda=H+\lambda f$ is affine in $\lambda$, its second
derivative vanishes. Unrolling the recursion from $\ell=r$ down to
$\ell=1$ therefore yields
\begin{align}
\frac{\d^2}{\d\lambda^2}P_0(H+\lambda f)\Big|_{\lambda=0}
&=
\sum_{\ell=1}^{r}\zeta_\ell
\Big\langle\mathrm{Var}_{<\ell}(f)\Big\rangle_{\ell-1}.
\label{eq:secondvar}
\end{align}
This proves \eqref{eq:hessian} for $g=f$. The general bilinear identity \eqref{eq:hessian} follows by
polarisation. 
\end{proof}

As a corollary of Proposition \ref{lem:pressure}, one can directly derive the linear response of a system described by a multiscale measure: 

\begin{corollary}[Linear Response of a multiscale measure]\label{cor:response}
Let $H\to H+\lambda f$ and let $\langle\,\cdot\,\rangle_\lambda$ denote the
corresponding multiscale average. Then, for every observable $O$  one has 
\be\label{eq:response}
\frac{\d}{\d\lambda}\big\langle O\big\rangle_\lambda\Big|_{\lambda=0}
=\sum_{\ell=1}^{r}\zeta_\ell\,\Big\langle\,\mathrm{Cov}_{<\ell}(O,f)\,\Big\rangle_{\ell-1}\,.
\ee
Moreover:
\begin{enumerate}
\item[(a)]  at fixed $(x_{\ell-1},\dots,x_1)$ the conditional measure $p^{<\ell}$
obeys the standard fluctuation--response relation at its own scale parameter,
\be
\frac{\d}{\d\lambda}\big\langle O\big\rangle_{<\ell,\lambda}\Big|_{\lambda=0}
=\zeta_\ell\,\big (\langle O\,f_{\ell}\big \rangle_{<\ell}-\langle O \rangle_{<\ell} \langle f_{\ell} \rangle_{<\ell} \big);
\ee
\item[(b)] if $f$ depends on $x_1$ only, then \eqref{eq:response} collapses to the
single term $\zeta_1\,\mathrm{Cov}(O,f)$;
\item[(c)] if all the $\zeta_\ell$ coincide, \eqref{eq:response} collapses to
$\zeta\,\mathrm{Cov}(O,f)$.
\end{enumerate}
\end{corollary}

\subsection{Examples of multiscale measures in Statistical Mechanics}
\label{sec:related-models}
This paragraph aims to show that the multiscale measure is a cornerstone of the theory of mean field spin glasses, highlighting the instances where it has appeared, and giving a possible dynamical interpretation for their origin. We first note that 
Definition~\ref{def:multiscale measure} was introduced in general abstract terms  for finite alphabets,  but the recursion \eqref{eq:recupartion} makes sense verbatim when each $X_\ell$ carries an a priori probability  measure (e.g.\ Gaussian) and the sums are replaced by the corresponding expectations $\E_\ell$ (the same convention is used in Section~\ref{sec:chinese}). 

\paragraph{Guerra's RSB bound of the Parisi solution.}
In this context, multiscale measures first appeared in Guerra's broken replica symmetry bound \cite{Guerra2002BrokenRS} of the Parisi solution of the Sherrington-Kirkpatrick (SK) model \cite{MPV}. The main idea is to compare the interacting spin system with a system of $N$ independent spins in a random field with a hierarchical covariance structure induced  by a multiscale measure. More precisely the system used to mimic the SK model has a pressure $P_0= \log Z_0$ obtained recursively as in \eqref{eq:genfunc},  starting with
\be
Z_{r+1}=\sum_{\sigma\in\{\pm1\}^N} \exp\left(\sum_{i\leq N}\sum_{\ell\leq r} g^{(\ell)}_i\sigma_i\right)
\ee
where $\{g^{(\ell)}_i,i\leq N\}$ are independent copies of a centered Gaussian r.v. At each level $\ell$ the corresponding family $g^{(\ell)}$ is averaged out as in \eqref{eq:rec press}, where summations are replaced by Gaussian expectations.
In this framework the Parisi order parameter, is related to the collection $\zeta$ of the a multiscale measure, each level of \textit{replica symmetry breaking} being a level of partial equilibrium. Dynamical aspects of this construction are discussed in \cite{Contucci_2019,Contucci_2021,alberici2024convergence}.\\

\paragraph{The multiscale Sherrington--Kirkpatrick model.}
In the Guerra's scheme  the multiscale structure is used as an artificial device to bound the pressure of the SK model. In   \cite{MultiscaleSK_CM2018} the authors consider the multiscale structure as a built-in feature of the random interaction of   the model itself. The pressure of the  model is obtained by a recursion of the type \eqref{eq:genfunc} with starting point 
 
\be
Z_{r+1}=\sum_{\sigma\in\{\pm1\}^N} \exp\left(\sum_{i,j\leq N}\sum_{\ell\leq r} g^{(\ell)}_{ij}\sigma_i\sigma_j\right)
\ee
where $\{g^{(\ell)}_{i,j},i\leq N\}$ are independent copies of a centered Gaussian r.v. with some $\ell$-dependent variance value. The random couplings of the SK model are split into $r$ independent Gaussian families. Here in addition to spins variables  each family $g^{(\ell)}$ is a thermodynamic degree of freedom with its own temperature: the spins thermalise first, then $g^{(r)}$, then $g^{(r-1)}$ and so on.  The pressure per particle converges to a Parisi-type variational formula \cite{MultiscaleSK_CM2018} where the presence of the apriori multiscale structure is reflected on the phases of the model \cite {camilli2025phase}. Here the scale parameters $\zeta$ (temperatures of the different thermal baths) become pinned values of the Parisi order parameter.

\paragraph{Inference and learning.}
In inference and learning, the hierarchy of scales is informational rather than temporal, but the underlying measure is the same. In the maximum-entropy treatment of a system in contact with a large but finite bath \cite{Pre13}, and more generally of composite systems \cite{Ram22}, the constraints fix the entropy of a subsystem conditionally on the rest, leading to the variational problem of Section~\ref{sec:varprinc}. A similar structure appears in transfer learning \cite{survey_transfer,Gerace_2022,TransferPRLFede}, where a representation (learned weights) learned from a \enquote{source} training dataset $\mathcal D_s$ is reused and adapted to a \enquote{target} training dataset $\mathcal D_t$. In broad terms, some weights of a machine can be trained for a certain task and then only fine-tuned in a second moment to perform another task.

More specifically, let $W_s$ denote the weights learned from the source task and let $(a,W_t)$ denote the parameters trained on the target task. We equip these parameters with some apriori measure (non-necessarily discrete), denoted as $P_a,P_W$. At fixed datasets, define
\begin{align}
Z_1(W_s)
&=\int dP_a(a),dP_W(W_t),
\exp\left(
-\beta\mathcal L_t(a,W_t;\mathcal D_t)
-\frac{\kappa}{2}\left|W_s-W_t\right|_2^2
\right),    \nonumber\\
Z_0^\zeta
&=\mathbb E_{W_s\mid\mathcal D_s}
\left[Z_1(W_s)^\zeta\right].
\end{align}
For machine learning studies, accessing the logarithm of $Z_0$ is crucial to estimate the generalization capabilities of such machines.

The above is the recursion \eqref{eq:recupartion} for a two-level multiscale measure: the target parameters $(a,W_t)$ are the fast variables in $X_2$ and are integrated first, with $\zeta_2=1$, while the source weights $W_s$ are the slow variables in $X_1$ and are integrated afterwards, with $\zeta_1=\zeta$. The parameter $\kappa$ controls the strength of the transfer: for $\kappa=0$ the target representation is trained independently of the source one, whereas $\kappa\to\infty$ imposes $W_t=W_s$, corresponding to a frozen transferred representation. The scale parameter $\zeta$, instead, distinguishes the learning protocol: in the limit $\zeta\to0$ one recovers complete transfer learning, where the source model is learned first and then treated as completely quenched during training on the target task; $\zeta=1$ corresponds to multitask learning, where the source and target tasks are learned simultaneously.

\paragraph{Multiscale measures from dynamics.} A multiscale measure does not describe a genuine thermal equilibrium, but rather an \emph{adiabatic steady state} of a dynamics coupled to multiple thermal baths and evolving on widely separated time scales (see e.g.~\cite{Kurchan-Cugliandolo,leuzzi2007thermodynamics,Vulpiani17,Contucci_2021}). For continuous variables, as an instance, consider the overdamped Langevin dynamics
\be
\lambda_\ell\,\d x_\ell(t)
=
-\nabla_\ell V(\bx(t))\,\d t
+\sqrt{\frac{2\lambda_\ell}{\beta_\ell}}\,\d W_\ell(t),
\qquad
0<\lambda_r\ll\lambda_{r-1}\ll\cdots\ll\lambda_1,
\ee
where the $W_\ell$ are independent Wiener processes. In the limit of infinitely separated time scales, $x_r$ first reaches its conditional Gibbs equilibrium while the slower variables remain frozen. Its averaged force derives from the corresponding free energy, which becomes the effective potential for $x_{r-1}$; iterating this adiabatic elimination yields precisely the recursion \eqref{eq:recupartion}, with $H=-\beta_rV$ and $\zeta_\ell=\beta_\ell/\beta_r$. For two scales, convergence to this non-equilibrium stationary measure has been established rigorously in \cite{alberici2024convergence}. An analogous recursive mechanism holds for finite-state continuous-time Markov chains with separated transition rates \cite{AlbericiGabrielliPallotta2026}.

\section{Multiscale measures from variational principles}\label{sec:varprinc}
The purpose of this section is to identify a variational problem whose solution is the multiscale measure \eqref{def:multiscale measure}. We do so through the Maximum Entropy (MaxEnt) principle \cite{Jaynes} under suitable constraints. The latter cannot be linear in the probability weights, since this would produce a standard Boltzmann-Gibbs measure. Hereby we show that the appropriate constraints fix the conditional entropy at the various levels. 

The Shannon entropy of a probability measure $\pi$ over a discrete space $\Omega$ is defined as
\begin{align}
    \mathcal{S}_\Omega[\pi]=-\sum_{\omega\in\Omega}\pi(\omega)\log\pi(\omega)
\end{align}where $\log$ can be in various bases. In this paper we consider the natural logarithm. In the following we shall omit the subscript of $\mathcal{S}$.

Keeping in mind the decomposition \eqref{eq:probdecomposition}, and using basic Shannon entropy properties, one can prove the following decomposition, known as chain rule for conditional entropies: 
\be\label{eq:entdecompo}
\mathcal{S}[p]=\sum_{1\leq\ell\leq r} 
\,\mathcal{S}^{\ell}[p]\,,\quad\quad
\mathcal{S}^{\ell}[p]:=\Big\langle\,\mathcal{S}[p^{<\ell}]\,\Big\rangle_{\ell-1}\,.
\ee
Hereby we put forward a variational principle whose solution is given by the multiscale measure:

\begin{proposition}\label{prop:var_princ_Maxent} Let $X$ be as above and denote the set of probability distribution on $X$ by $\mathcal{P}_X$. Let $\mu$ and $(\gamma_\ell)_{2\leq\ell\leq r}$ be collections of real parameters satisfying
    \be\label{eq:positivity-hypothesis}
    \mu>0,\quad  1+\gamma_\ell>0\qquad\text{for every }\quad 2\leq \ell\leq r.
    \ee
    Consider the variational problem
    \begin{align}\label{eq:phi_def}
        \sup_{p\in\mathcal{P}_X} \phi[p],\quad \phi[p]:=\mathcal{S}[p]+\mu\langle H\rangle+\sum_{\ell=2}^r\gamma_{\ell}\,\mathcal{S}^\ell[p]\,.
    \end{align}
    
    The functional $\phi$ is strictly concave  on $\mathcal{P}_X$ and the unique solution of \eqref{eq:phi_def}
    is the $r$-level multiscale measure of Definition~\ref{def:multiscale measure} with scale parameters
    \be\label{eq:zeta-dictionary}
    \zeta_\ell=\frac{\mu}{1+\gamma_\ell}\,, \quad 1\leq\ell\leq r\,,\qquad \gamma_1:=0.
    \ee
    Furthermore, $\phi$  evaluated at the optimal distribution equals $\mu \log Z_0$.
\end{proposition}

\begin{remark}
The parameters $\mu$ and $(\gamma_\ell)_{2\leq\ell\leq r}$ are to be
interpreted as Lagrange multipliers fixing the internal energy and the conditional entropies at the various scales. For example, when $r=2$,
fixing the internal energy and one of the conditional entropies leads to the maximization of $\phi=\mathcal{S}^1+(1+\gamma_2)\mathcal{S}^2+\mu\langle H\rangle$,
and hence to a two-level multiscale measure. The two scales are genuinely distinct when $\gamma_2\neq0$, whereas for $\gamma_2=0$ the
usual Gibbs measure is recovered. Indeed, the crucial point is that fixing a conditional entropy can introduce an intrinsic asymmetry
between the levels, namely $p^{<1}$ and $p^{<2}$, which is absent from the standard Gibbs variational principle.
\end{remark}

We first need the following Lemma, whose proof is a direct consequence of the chain rule for the total entropy, and of the fact that the single $S^{\ell}$'s are concave.

\begin{lemma}[Concavity of the imbalanced functional]\label{lem:concavity}
For every $\ell\leq r$ the conditional entropy $\mathcal{S}^{\ell}[p]$ defined in
\eqref{eq:entdecompo} is a concave function of the joint law $p$ on $X$.
Consequently, for every choice of coefficients $c_1,\dots,c_r>0$, every
$\mu\in\R$ and every $H:X\to\R$, the functional
\be\label{eq:phi-general}
\phi[p]=\sum_{\ell=1}^{r}c_\ell\,\mathcal{S}^{\ell}[p]+\mu\,\langle H\rangle_p
\ee
is strictly concave on the simplex of probability measures on $X$.
\end{lemma}

\begin{proof}[Proof of Proposition \ref{prop:var_princ_Maxent}]
Using \eqref{eq:entdecompo}, the variational problem we need to solve can be recast as
\be\label{eq:varprincipmany}
\phi[p]=\mathcal{S}^1[p]+\sum_{2\leq \ell\leq r}(1+\gamma_{\ell})\mathcal{S}^{\ell}[p]+\mu \big\langle\,H\,\big\rangle\,.
\ee
Notice that for any $\ell\leq r$ if we fix the marginals $p^{(\ell-1)}$ then $\mathcal{S}^{\ell}$ depends only on the choice of $p^{<\ell}$. Hence one can maximize $\phi$ in a hierarchical way starting from $p^{<r}$, fixing $p^{(r-1)}$. Namely we write 
\be\label{eq:phideco}
\phi[p]= \mathcal{S}^1[p]+\sum_{2\leq \ell\leq r-1}(1+\gamma_{\ell})\mathcal{S}^{\ell}[p]+ \big\langle \varphi_r\big\rangle_{r-1}
\ee
where 
\be\label{functerre}
\varphi_r=(1+\gamma_r)\mathcal{S}[p^{<r}]+\mu\big\langle\,H\,\big\rangle_{<r}
\ee
and maximize $\varphi_r$  w.r.t. the choice of $p^{<r}$. This is the standard maximization procedure under the linear constraint for the internal energy, which this time is conditional on the realization of the other degrees of freedom $x_{r-1},\dots,x_1$ that will thermalize later. The $\sup$ is attained at
\be
p^{<r}\equiv p^{<r}(x_r|x_{r-1}\ldots x_1)=\dfrac{1}{\mathcal Z_{r-1}}e^{\frac{\mu}{1+\gamma_r}H(\bx)}\,,\quad\, \mathcal{Z}_{r-1}=Z^{\frac{\mu}{1+\gamma_r}}_{r-1}=\sum_{x_r}e^{\frac{\mu}{1+\gamma_r}H(\bx)}
\ee
substituting the optimal conditional distribution back into the functional \eqref{functerre} yields 
\be
\varphi_r=(1+\gamma_r)\log \mathcal{Z}_{r-1}=\mu \log Z_{r-1}\,.
\ee
In the next step we observe that  $p^{(r-1)}=p^{<r-1}p^{(r-2)}$ and then
\be
\phi[p]= \mathcal{S}^1[p]+\sum_{2\leq \ell\leq r-2}(1+\gamma_{\ell})\mathcal{S}^{\ell}[p]+ \big\langle \varphi_{r-1}\big\rangle_{r-2}
\ee
 where 
 \be
 \varphi_{r-1}=(1+\gamma_{r-1})\mathcal{S}[p^{<r-1}]+\mu\big\langle\log Z_{r-1} \big\rangle_{<r-1}
\ee
The usual maximization procedure yields
\be
p^{<r-1}(x_{r-1}|x_{r-2},\ldots ,x_1)\propto e^{\frac{\mu}{1+\gamma_{r-1}}\log Z_{r-1}}\,.
\ee
The same reasoning can be iterated until the end, namely the maximization over $p^{<1}$ yielding the multiscale measure, with the scale parameters $\bzeta$ given exactly by \eqref{eq:zeta-dictionary}. Iterating the corresponding optimal values yields $\phi[p]=\mu\log Z_0$ at the maximizer.
\end{proof}

\subsection{Two temperature thermodynamics} \label{sec:twotemperatures}
We now  discuss  some thermodynamic aspects  of the solution of the variational problem \eqref{eq:phi_def}. We focus on the case $r=2$, hence we look for a probability
measure $p=\left(p(x_2,x_1)\right)_{(x_2,x_1)\in X}$ on a product space $X=X_2\times X_1$
such that
\be\label{eq:vaprob2}
\max_{p}\,\mathcal{S}[p]\quad \text{with}\quad\big\langle\,H\,\big\rangle = \,E\quad
\text{ and}\quad \mathcal{S}^{2}[p]=\Big\langle\,\mathcal{S}[p^{<2}]\,\Big\rangle_1=S_2\,.
\ee
Proposition~\ref{prop:var_princ_Maxent} with $r=2$, $\gamma_1=0$, $\gamma_2=\gamma$,
the scale parameters being $\zeta_2=\mu/(1+\gamma)$ and $\zeta_1=\mu$, gives:
\be\label{eq:var2result}
\begin{aligned}
p^{<2}&\equiv p^{<2}(x_2|x_1)=
  \dfrac{e^{\frac{\mu}{1+\gamma}H(x_2,x_1)}}{\mathcal{Z}_1(x_1)},
  \quad \mathcal{Z}_1(x_1)=\sum_{x_2}e^{\frac{\mu}{1+\gamma}H(x_2,x_1)}=Z_1^{\frac{\mu}{1+\gamma}},
\\
p^{<1}&\equiv p^{<1}(x_1)=\left(\dfrac{Z_1}{Z_0}\right)^{\mu}.
\end{aligned}
\ee

We claim that the measure \eqref{eq:var2result} is related to a
two-temperature thermodynamics \cite{leuzzi2007thermodynamics} where the
degrees of freedom $x_1$ and $x_2$ are in contact with thermal reservoirs
at inverse temperatures $\beta_1$ and $\beta_2$ respectively. We recall that, in the convention of Definition~\ref{def:multiscale measure}, the Boltzmann factor is $e^{\zeta H}$: $H$ plays the role of minus the energy, so that the standard entropy--energy relation reads $\partial\mathcal{S}/\partial E=-\beta$, and inverse temperatures are measured in units of $\beta$ (see Remark~\ref{rem:physconv}). The Lagrange
multipliers $\mu$ and $\gamma$ can be expressed in terms of $E$ and $S_2$
observing that by  Proposition~\ref{prop:var_princ_Maxent},
the function $(\mu,\gamma)\;\longmapsto\;\mu P_0(\mu,\gamma)=\sup_p\phi[p]$  is \emph{convex} in $(\mu,\gamma)$, being a supremum of affine functions. The optimizer $p^*$ satisfies

\be\label{eq:envelope}
\frac{\partial\bigl(\mu P_0\bigr)}{\partial\mu}
  =\Big\langle H\Big\rangle_{p^*}=E\,,
\qquad
\frac{\partial\bigl(\mu P_0\bigr)}{\partial\gamma}
  =\mathcal{S}^2[p^*]=S_2\,.
\ee
Therefore, $E$ and $S_2$ are the \emph{conjugate variables}
of the convex potential $\mu P_0$ with respect to $\mu$ and $\gamma$
respectively. Since
$\mu P_0$ is convex,  its Legendre--Fenchel conjugate is
\be\label{eq:legendre}
\Phi(E,S_2)
  :=\sup_{\substack{\mu>0\\1+\gamma>0}}
    \Big(\mu E+(1+\gamma)S_2-\mu P_0(\mu,\gamma)\Big)
\ee
and it is \emph{convex} in $(E,S_2)$, equivalently  

\be\label{eq:Phi-meaning}
-\Phi(E,S_2)=\max\Big\{\mathcal{S}^1[p]\;:\;\langle H\rangle_p=E,\ \mathcal{S}^2[p]=S_2\Big\}\,.
\ee
Hence $-\Phi(E,S_2)$ is the maximal entropy of the $x_1$ degrees of freedom compatible with the constraints \eqref{eq:vaprob2}. The supremum in \eqref{eq:legendre} is attained at the unique
$(\mu^*,\gamma^*)$ satisfying \eqref{eq:envelope}, and the envelope theorem gives
\be\label{eq:multipliers}
\frac{\partial\Phi}{\partial E}=\mu^*\,,
\qquad
\frac{\partial\Phi}{\partial S_2}=1+\gamma^*\,.
\ee

Since $S_2$ represents the entropy of the $x_2$ degrees of freedom given
$x_1$,  the inverse temperature $\beta_2$ is identified through the
standard entropy-energy relation. A level curve of $\Phi$ is, by \eqref{eq:Phi-meaning}, a curve of constant entropy of the $x_1$ degrees of freedom: along it, a variation of the energy is entirely absorbed by the $x_2$ degrees of freedom. From \eqref{eq:multipliers}, by the implicit function
theorem:
\be\label{eq:firstbeta}
\frac{\partial S_2}{\partial E}\Big|_{\Phi}
  =-\frac{\partial\Phi/\partial E}{\partial\Phi/\partial S_2}
  =-\frac{\mu^*}{1+\gamma^*}
  =-\beta_2\,,
\ee
the sign being the one dictated by the convention $\partial\mathcal{S}/\partial E=-\beta$ recalled above. Hence $\beta_2=\mu^*/(1+\gamma^*)=\zeta_2$.

For the $x_1$ degrees of freedom, with $\zeta_1=\mu^*$, the variational
principle \eqref{eq:phi_def} reduces (see the proof of Proposition~\ref{prop:var_princ_Maxent}) to
\be\label{eq:varp1}
\max_{p^{<1}}\Big(\mathcal{S}[p^{<1}]
  +\mu^*\langle\log Z_1(x_1)\rangle_1\Big)\,.
\ee
Setting $\tilde H(x_1):=-\frac{1}{\beta_2}\log\mathcal{Z}_1(x_1)$ as the
effective Hamiltonian (the $x_2$ free energy conditioned on $x_1$) and
$\tilde{\mathcal{S}}:=\mathcal{S}[p^{<1}]$, and using
$\log Z_1=\frac{1}{\zeta_2}\log\mathcal{Z}_1=\frac{1}{\beta_2}\log\mathcal{Z}_1=-\tilde H$, the objective becomes
\be\label{eq:varp1clean}
\tilde{\mathcal{S}}-\mu^*\langle\tilde H\rangle_1\,.
\ee
This is the standard Gibbs variational principle for the $x_1$ subsystem with energy $\tilde H$ (here with the usual sign) at thermal
equilibrium, whose entropy-energy relation gives
\be\label{eq:secondbeta}
\frac{\partial\tilde{\mathcal{S}}}{\partial\langle\tilde H\rangle}
  =\beta_1=\mu^*
\,.
\ee
The two scale parameters therefore coincide
with the two inverse temperatures:
\be\label{eq:betadict}
\zeta_1=\mu^*=\beta_1\,,\qquad
\zeta_2=\frac{\mu^*}{1+\gamma^*}=\beta_2\,,\qquad 1+\gamma^*=\frac{\beta_1}{\beta_2}\,.
\ee

We also emphasize that by Corollary \ref{cor:response} and with  the identification \eqref{eq:betadict}, the partial thermal equilibrium picture  occurs also in the sense of linear
response theory.\\ 

Let us now briefly discuss the effect of the constraint $S_2$ on the conditional entropy in some
limiting cases. By the basic properties of Shannon entropy,
\be\label{eq:entropybounds}
0\leq\Big\langle\mathcal{S}[p^{<2}]\Big\rangle_1\leq\log|X_2|\,.
\ee
$S_2=0$ occurs when $p^{<2}(\cdot|x_1)$ is concentrated on a single atom:
\be
p^{<2}(x_2|x_1)=\delta_{x_2,x^*(x_1)}
\ee
for some $x^*(x_1)\in X_2$ depending on $x_1$. Imposing such concentration, the
variational problem~\eqref{eq:vaprob2} becomes
\be\label{eq:limitproblem}
\begin{cases}
\max_{p^{<1},\,x^*}\;\mathcal{S}[p^{<1}]\\[4pt]
\displaystyle\sum_{x_1}p^{<1}(x_1)\,H(x^*(x_1),x_1)=E
\end{cases}.
\ee
The optimization over $x^*$ forces $x^*(x_1)$ to be the \emph{maximizer} of
$x_2\mapsto H(x_2,x_1)$ for each fixed $x_1$, consistently with the limit
$\beta_2\to+\infty$ selecting the maximum-energy $x_2$ configuration.
In the two-temperature framework this scenario corresponds to $\beta_2\to+\infty$
with $\beta_1<\infty$.

In other words, $p^{<1}(x_1)$ can be viewed as the measure on the maximizer(s) of
$x_1\mapsto\max_{x_2}H(x_2,x_1)$, assigning to $x_1$ a weight proportional to
$e^{\beta_1 H(x^*(x_1),x_1)}$. In particular, if $E=\max_{x_2,x_1}H(x_2,x_1)$ then
$\beta_1\to\infty$ as well and all mass concentrates on the global maximizer(s) of $H$.
Conversely, for $S_2=\log|X_2|$ the only admissible choice is
$p^{<2}(\cdot|x_1)=1/|X_2|$, i.e.\ the uniform measure on $X_2$, which occurs
when $\beta_2\to 0$.

\section{A large deviation perspective}
\label{sec:large-deviation-perspective}

We recall the classical Boltzmann argument on the asymptotics of a multinomial
distribution, which can be regarded as one of the starting points of statistical
mechanics and large deviation theory \cite{Ellis99}. Let $X$ be a finite set and
let $q=(q_x)_{x\in X}$ be a probability measure with full support. If $n$ balls
are independently assigned to the boxes indexed by $X$ according to $q$, then
the vector $N=(N_x)_{x\in X}$ of occupation numbers follows a \textit{multinomial distribution}
$\mathcal M(n,q)$, and $\widehat p_n:=N/n$ is the corresponding empirical
probability measure. A standard   estimate (see, for example,
\cite[Lemma~2.1.9]{DemboZeitouni1998}) is the following:

\begin{lemma}
\label{lem:multinomial-method-of-types}
Let $X$ be a finite set, let $n\in\mathbb N$, and let $q=(q_x)_{x\in X}\in \mathcal{P}_X$ be a
probability measure on $X$ with full support. For every
$p\in\mathcal P_X\cap n^{-1}\mathbb Z^X$,
\be
(n+1)^{-|X|}e^{-nD_{\mathrm{KL}}(p\|q)}
\leq
\mathbb P\bigl(\mathcal M(n,q)=np\bigr)
\leq
e^{-nD_{\mathrm{KL}}(p\|q)}.
\label{eq:multinomial-method-of-types}
\ee
Consequently,
\be
\frac1n\log\mathbb P\bigl(\mathcal M(n,q)=np\bigr)
=-D_{\mathrm{KL}}(p\|q)+O\left(\frac{\log n}{n}\right)
\label{eq:multinomial-logarithmic-asymptotic}
\ee
uniformly over all $p\in\mathcal P_X\cap n^{-1}\mathbb Z^X$.
\end{lemma}

The empirical measure  $\widehat p_n$ therefore satisfies Sanov's large deviation principle
with rate function $D_{\mathrm{KL}}(\cdot\|q)$. When $q$ is uniform,
$D_{\mathrm{KL}}(p\|q)=\log|X|-S[p]$, so minimizing the rate function under an
energy constraint reproduces the Gibbs variational principle.\\

We aim to construct
a random experiment whose rate function is precisely the entropic part of the functional \eqref{eq:phi_def} in Proposition~\ref{prop:var_princ_Maxent}.
\subsection{The case of two scales}
\label{subsec:two-scale-reinforcement}
In order to present the idea let's start with the case $r=2$. Let $X=X_2\times X_1$ and let $q$ be a  probability distribution on $X$ with full support. One can imagine to have boxes indexed by $X_1$, that we call \emph{parent boxes}, and inside each of them there are $|X_2|$ \emph{child boxes}. Hence a specific child box is indexed by an ordered pair $(i,j)\in X=X_2\times X_1$ ($i$-th child of the $j$-th parent).  At the first level, $n$
balls are independently assigned to parent boxes according to $q^{<1}$, so
\be
N^{(1)}\sim\mathcal M(n,q^{<1}).
\label{eq:two-scale-first-level}
\ee
Fix $\gamma>-1$, after entering a parent box $j\in X_1$, its population is
rescaled deterministically to
\be
T_j:=\left\lfloor(1+\gamma)N_j^{(1)}\right\rfloor.
\label{eq:two-scale-reinforcement}
\ee
Conditionally on $N^{(1)}$, and independently across $j\in X_1$,
\be
\bigl(N_{i,j}^{(2)}\bigr)_{i\in X_2}
\sim
\mathcal M\bigl(T_j,q^{<2}(\cdot\mid j)\bigr).
\label{eq:two-scale-second-level}
\ee
Define
\be
\widehat p_n^{<1}(j):=\frac{N_j^{(1)}}n\,,\quad \widehat p_n^{<2}(i\mid j):=\begin{cases}\frac{N_{i,j}^{(2)}}{T_j} \quad \text{if} \quad T_j>0\\
q^{<2}(\cdot\mid j)\quad\text{if}\quad  T_j=0.
\end{cases}
\label{eq:two-scale-second-kernel}
\ee

The cases for $\widehat p^{<2}$ are needed only to have the empirical distribution always well defined, even in the extremely unlikely event that there are no balls to distribute in the $j$-th parent box.
The empirical distribution of the process is then defined as
\be
\widehat p_n(i,j):=\widehat p_n^{<1}(j)\widehat p_n^{<2}(i\mid j).
\label{eq:two-scale-empirical-measure}
\ee

\begin{figure}
    \centering
    \includegraphics[width=0.5\linewidth]{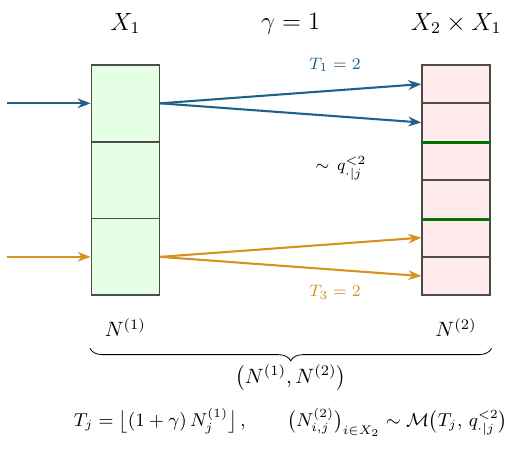}
    \caption{
    Schematic representation of a two--scale reinforced multinomial process.
    The space $X_{1}$ is partitioned into three equal parent cells; each parent cell contains two child cells. Hence, every child cell is labeled by a couple of indices in $X_{2}\times X_{1}$.
    Blue and yellow arrows represent balls cast into the parent (or child) cells. 
    }
    \label{fig:active}
\end{figure}

The general result proved below implies that $ \widehat p_n$ satisfies an Large Deviation Principle
with speed $n$ and rate function
\be
I_q(p)=D_{\mathrm{KL}}(p^{<1}\|q^{<1})
+(1+\gamma)\sum_{j\in X_1}p^{<1}(j)
D_{\mathrm{KL}}\left(p^{<2}(\cdot\mid j)\middle\|q^{<2}(\cdot\mid j)\right).
\label{eq:two-scale-rate-function}
\ee

\begin{remark}
Notice that the  reinforcement does not change the typical outcome, which remains $q$,
but changes the exponential cost of deviations of the conditional empirical
distribution at the second level. For $\gamma=0$ the ordinary multinomial rate
is recovered. For $-1<\gamma<0$ second-level deviations are penalized less, and
for $\gamma>0$ they are penalized more.

\end{remark}

\subsection{A hierarchical reinforcement process}
\label{subsec:hierarchical-reinforcement}

Consider now   the general case $r\geq1$. For $0\leq\ell\leq r$ set
\be
B_\ell:=X_\ell\times\cdots\times X_1,
\qquad B_0:=\{\varnothing\},
\qquad B_r=X.
\label{eq:hierarchical-boxes}
\ee
A box $b=(x_\ell,\ldots,x_1)\in B_\ell$ has children
$(x_{\ell+1},b)\in B_{\ell+1}$. Let $q$ be a probability measure on $X$ with
full support.

\begin{definition}[Reinforcement process]
\label{def:scheme}
Fix $m_2,\ldots,m_r>0$, set $m_1:=1$, and let $n\in\mathbb N$.
We construct the process recursively over the levels of the hierarchy. Set $N_{\varnothing}^{(0)}:=n$
and for $\ell=1,\ldots,r$, suppose that the populations
$\bigl(N_b^{(\ell-1)}\bigr)_{b\in B_{\ell-1}}$ have already been
constructed. For each $b\in B_{\ell-1}$, define
\be
T_b:=\left\lfloor m_\ell N_b^{(\ell-1)}\right\rfloor.
\label{eq:population-recursion}
\ee
Conditionally on the counts constructed up to level $\ell-1$, and
independently across $b\in B_{\ell-1}$, let
\be
\bigl(N_{(x,b)}^{(\ell)}\bigr)_{x\in X_\ell}
\sim
\mathcal M\bigl(T_b,q^{<\ell}(\cdot\mid b)\bigr).
\label{eq:allocation}
\ee
In particular, since $m_1=1$, one has $T_{\varnothing}=n$ at the first
level. The reinforcement process is the random array
\be
\boldsymbol N
:=
\left(N_b^{(\ell)}\right)_{1\leq\ell\leq r,\ b\in B_\ell}.
\label{eq:reinforcement-process}
\ee
\end{definition}

Note that, not all arrays of integers $\mathfrak n=(n_b^{(\ell)})^{\ell\leq r}_{b\in B_{\ell}}$ are \emph{realizable} in the sense of the above process, since there are constraints regarding the number of children at each level inherited by \eqref{eq:population-recursion}, \eqref{eq:allocation}. We call the realizable set
$\mathfrak N_n$. For $\mathfrak n\in\mathfrak N_n$ set
\be
t_b:=
\begin{cases}
n,&\ell=1,\ b=\varnothing,\\
\lfloor m_\ell n_b^{(\ell-1)}\rfloor,&2\leq\ell\leq r,
\end{cases}
\label{eq:array-parent-population}
\ee
and define a conditional probability
\be
p_{\mathfrak n}^{<\ell}(x\mid b):=
\begin{cases}
n_{(x,b)}^{(\ell)}/t_b,&t_b>0,\\
q^{<\ell}(x\mid b),&t_b=0
\end{cases}
\label{eq:array-empirical-kernels}
\ee
which is representing the normalized histogram associated to $\mathfrak n$. By the chain rule \eqref{eq:probdecomposition} the family $\{p_{\mathfrak n}^{<\ell}, \ell\leq r\}$ determines a probability measure $p_{\mathfrak n}$ on $X$. We denote its marginal on $B_\ell$ by $p_{\mathfrak n}^{(\ell)}$. We denote by
$\widehat p_n$ the random probability measure on $X$ associated to reinforcement process $\boldsymbol{N}$.

\begin{theorem}
\label{thm:reinforced-multinomial-ldp}
Let $q$ be a probability measure with full support on $X$, let
$\boldsymbol N$ be the reinforcement process of
Definition~\ref{def:scheme}, and let $\widehat p_n$  be its empirical measure on $X$.  Given  $p\in\mathcal P_X$, let $\mathfrak d_{n,p}\in\mathfrak N_n
$ be a sequence of
realizable arrays such that 
\be
\left\|
p_{\mathfrak d_{n,p}}-p
\right\|_1
=\sum_{x\in X}|p_{\mathfrak d_{n,p}}(x)-p(x)|=o(1)
\label{eq:generic-discretization-convergence}
\ee
as $n\to\infty$. Then
\be
\frac1n
\log
\mathbb P\left(
\boldsymbol N=\mathfrak d_{n,p}
\right)
=
-I_q(p)+o(1),
\label{eq:generic-point-probability}
\ee
where,  setting $
c_\ell:=\prod_{k=1}^\ell m_k$, 
\be
I_q(p)
:=
\sum_{\ell=1}^r
c_\ell
\sum_{b\in B_{\ell-1}}
p^{(\ell-1)}(b)
D_{\mathrm{KL}}
\left(
p^{<\ell}(\cdot\mid b)
\middle\|
q^{<\ell}(\cdot\mid b)
\right)
\label{eq:reinforced-rate-function}.
\ee

Moreover, $(\widehat p_n)_{n\geq1}$ satisfies a large deviation principle on
$\mathcal P_X$ with speed $n$ and rate function $I_q$ which  is continuous on $\mathcal{P}_X$, it is a good rate function and has a unique zero, attained at $p=q$.
\end{theorem}

Using Varadhan's Lemma \cite{DemboZeitouni1998}, we can link the result of Theorem \ref{thm:reinforced-multinomial-ldp} to Proposition \ref{prop:var_princ_Maxent}. In fact, by taking $q$ uniform, for large $n$, the scaled cumulant generating function of the energy $H$ w.r.t. to $\widehat p_n$ converges to the potential $\phi$ in  \eqref{eq:phi_def}.

\begin{remark}
\label{rem:largest-remainder-discretization}
The theorem  requires that, for every
$p\in\mathcal P_X$, one can choose a sequence of realizable arrays
$\mathfrak d_{n,p}\in\mathfrak N_n$ such that $p_{\mathfrak d_{n,p}}
\longrightarrow p$. Such arrays can be constructed explicitly by a \enquote{largest-remainder rule}. Recall that,
given an integer $t\geq0$ and a probability vector $u=(u_x)_{x\in Y}$
on a finite set $Y$, the largest-remainder rule produces integers
$(k_x)_{x\in Y}$ such that
\be
\sum_{x\in Y}k_x=t,
\qquad
\left|\frac{k_x}{t}-u_x\right|<\frac{1}{t}
\quad\text{for every }x\in Y.
\label{eq:largest-remainder-property}
\ee
Indeed, one first sets $k_x=\lfloor tu_x\rfloor$ and then distributes
the remaining units among the coordinates having the largest fractional
parts.

Using this method in our setting, recursively for each level $\ell$, one can readily get the uniform control
\be
\sup_{p\in\mathcal P_X}
\left\|
p_{\mathfrak d_{n,p}}-p
\right\|_1
\leq
\frac{C}{n}
\label{eq:largest-remainder-uniform-rate}
\ee
for a constant $C$ depending only on the fixed hierarchy and the
reinforcement parameters. The largest-remainder construction implies also the
stronger local estimate
\be
\frac1n
\log
\mathbb P\left(
\boldsymbol N=\mathfrak d_{n,p}
\right)
=
-I_q(p)
+
O\left(\frac{\log n}{n}\right)
\label{eq:largest-remainder-sharp-point-estimate}
\ee
uniformly over every compact set $p\in \mathcal P_X$ on which all
parent marginals are bounded away from zero.
\end{remark}

\begin{proof}[Proof of Theorem \ref{thm:reinforced-multinomial-ldp}]

We start with some preliminary estimates. At first we need a uniform control on the difference between the realizable populations arrays $n^{\ell}_b$ and reconstructed marginals $p^{(\ell)}_{\mathfrak{n}}$ over $B_\ell$.

\begin{lemma}
\label{lem:population-approximation}
Set
\be
A_1:=0,
\qquad
A_\ell:=1+m_\ell A_{\ell-1},
\qquad
2\leq\ell\leq r.
\label{eq:population-approximation-constants}
\ee
Then, for every $n\in\mathbb N$, every realizable array
$\mathfrak n\in\mathfrak N_n$, and every $1\leq\ell\leq r$, one has
\be
\left|
n_b^{(\ell)}
-
nc_\ell p_{\mathfrak n}^{(\ell)}(b)
\right|
\leq
A_\ell,
\qquad b\in B_\ell,
\label{eq:population-marginal-approximation}
\ee
and
\be
\left|
t_b
-
nc_\ell p_{\mathfrak n}^{(\ell-1)}(b)
\right|
\leq
A_\ell,
\qquad b\in B_{\ell-1}.
\label{eq:parent-population-marginal-approximation}
\ee
In particular,
\be
\frac{n_b^{(\ell)}}{n}
=
c_\ell p_{\mathfrak n}^{(\ell)}(b)
+
O\left(\frac1n\right)
\label{eq:normalized-population-marginal-approximation}
\ee
uniformly over $\mathfrak n\in\mathfrak N_n$ and $b\in B_\ell$.
\end{lemma}

\begin{proof}
We prove \eqref{eq:population-marginal-approximation} and
\eqref{eq:parent-population-marginal-approximation} simultaneously by
induction over $\ell$. At level $\ell=1$, the only parent box is the root
$b=\varnothing$. Since $t_\varnothing=n$, $c_1=m_1=1$, and
$p_{\mathfrak n}^{(0)}(\varnothing)=1$, one has $
t_\varnothing
=
n
=
nc_1p_{\mathfrak n}^{(0)}(\varnothing).$ Moreover, at the first-level  
\be
n_{x_1}^{(1)}
=
nc_1p_{\mathfrak n}^{(1)}(x_1),
\label{eq:first-level-population-exact}
\ee
hence \eqref{eq:population-marginal-approximation} also holds at level one
with $A_1=0$. Assume now that
\eqref{eq:population-marginal-approximation} holds at level $\ell-1$, for
some $2\leq\ell\leq r$. Thus,
\be
\left|
n_b^{(\ell-1)}
-
nc_{\ell-1}p_{\mathfrak n}^{(\ell-1)}(b)
\right|
\leq
A_{\ell-1},
\qquad b\in B_{\ell-1}.
\label{eq:population-approximation-induction-hypothesis}
\ee
Fix $b\in B_{\ell-1}$. By the definition of $t_b$, using $
\left|
\lfloor z\rfloor-z
\right|
<1$ and since
$c_\ell=m_\ell c_{\ell-1}$, one  obtains
\be
\left|
t_b
-
nc_\ell p_{\mathfrak n}^{(\ell-1)}(b)
\right|\leq
\left|
t_b
-
m_\ell n_b^{(\ell-1)}
\right|
+
m_\ell
\left|
n_b^{(\ell-1)}
-
nc_{\ell-1}p_{\mathfrak n}^{(\ell-1)}(b)
\right|<
1+m_\ell A_{\ell-1}
\label{eq:parent-population-approximation-induction}
\ee
This proves \eqref{eq:parent-population-marginal-approximation} at level
$\ell$. It remains to prove \eqref{eq:population-marginal-approximation} at level
$\ell$. Fix $x\in X_\ell$ and $b\in B_{\ell-1}$. We distinguish two cases. Suppose first that $t_b>0$. By the definition \eqref{eq:array-empirical-kernels} of the conditional measure associated  with $\mathfrak n$, one has $
n_{(x,b)}^{(\ell)}
=
t_bp_{\mathfrak n}^{<\ell}(x\mid b)$ .
On the other hand, the marginals of the reconstructed measure satisfy
$p_{\mathfrak n}^{(\ell)}(x,b)
=
p_{\mathfrak n}^{(\ell-1)}(b)
p_{\mathfrak n}^{<\ell}(x\mid b)$
consequently,
\be
\left|
n_{(x,b)}^{(\ell)}
-
nc_\ell p_{\mathfrak n}^{(\ell)}(x,b)
\right|
=
p_{\mathfrak n}^{<\ell}(x\mid b)
\left|
t_b
-
nc_\ell p_{\mathfrak n}^{(\ell-1)}(b)
\right|
\leq
A_\ell.
\label{eq:positive-parent-child-approximation}
\ee
Suppose instead that $t_b=0$, since the array is realizable, a parent box
with zero population produces no children, hence $n_{(x,b)}^{(\ell)}=0,
\, x\in X_\ell$.
Moreover,
\eqref{eq:parent-population-marginal-approximation} gives
\be
nc_\ell p_{\mathfrak n}^{(\ell-1)}(b)
\leq
A_\ell.
\label{eq:zero-parent-reconstructed-mass-bound}
\ee
Since  $t_b=0$, by convention, $
p_{\mathfrak n}^{<\ell}(\cdot\mid b)
=
q^{<\ell}(\cdot\mid b)$ 
and then using the marginal recursion, we obtain
\be
\left|
n_{(x,b)}^{(\ell)}
-
nc_\ell p_{\mathfrak n}^{(\ell)}(x,b)
\right|
=
nc_\ell p_{\mathfrak n}^{(\ell-1)}(b)
q^{<\ell}(x\mid b)
\leq
A_\ell.
\label{eq:zero-parent-child-approximation}
\ee

Thus \eqref{eq:population-marginal-approximation} holds at level $\ell$ and 
the induction is complete. Finally, dividing \eqref{eq:population-marginal-approximation} by $n$ gives
\be
\left|
\frac{n_b^{(\ell)}}n
-
c_\ell p_{\mathfrak n}^{(\ell)}(b)
\right|
\leq
\frac{A_\ell}{n}.
\label{eq:normalized-population-explicit-error}
\ee
Since the constants $A_\ell$ depend only on the fixed reinforcement
parameters, the estimate is uniform over
$\mathfrak n\in\mathfrak N_n$ and $b\in B_\ell$.
\end{proof}

The next lemma provides a point estimate of probability of a realizable array at the $n^{-1}\log$ scale.

\begin{lemma}
\label{lem:array-probability}
Let $q$ have full support on $X$. Then, uniformly over all realizable arrays
$\mathfrak n\in\mathfrak N_n$,
\be
\log
\mathbb P\left(
\boldsymbol N=\mathfrak n
\right)
=
-nI_q(p_{\mathfrak n})
+
O\bigl(\log(n+1)\bigr),
\label{eq:array-probability}
\ee

\end{lemma}

\begin{proof}
Fix $n\in\mathbb N$ and a realizable array $\mathfrak n
\in\mathfrak N_n$.
Keep in mind the definitions of $t_b$ and $p_{\mathfrak n}$ in  \eqref{eq:array-parent-population}
and \eqref{eq:array-empirical-kernels}
and, for brevity, set $
\nu_b^{(\ell)}
:=
p_{\mathfrak n}^{<\ell}(\cdot\mid b)$. By the recursive definition of the reinforcement process, conditionally on
the populations at level $\ell-1$, the multinomial allocations associated
with distinct boxes $b\in B_{\ell-1}$ are independent. The chain rule for
conditional probabilities therefore gives
\be
\begin{split}
\mathbb P\left(
\boldsymbol N=\mathfrak n
\right)
&=
\prod_{\ell=1}^r
\prod_{b\in B_{\ell-1}}
\mathbb P\left(
\mathcal M
\left(
t_b,q^{<\ell}(\cdot\mid b)
\right)
=
\left(
n_{(x,b)}^{(\ell)}
\right)_{x\in X_\ell}
\right).
\end{split}
\label{eq:realizable-array-factorization}
\ee

Notice that if $t_b=0$,
the corresponding count vector is identically zero and
\be
\mathbb P\left(
\mathcal M
\left(
0,q^{<\ell}(\cdot\mid b)
\right)
=
0
\right)
=
1.
\label{eq:zero-population-multinomial-factor}
\ee
Hence the factors with $t_b=0$ make no contribution to the logarithm of
\eqref{eq:realizable-array-factorization}. If $t_b>0$, then $\left(
n_{(x,b)}^{(\ell)}
\right)_{x\in X_\ell}
=
t_b\nu_b^{(\ell)}$ 
and $\nu_b^{(\ell)}$ is a probability vector in $\mathcal{P}_{ X_{\ell}}\cap\frac{1}{t_b}\mathbb{Z}^{X_{\ell}}$. Hence we can apply  Lemma~\ref{lem:multinomial-method-of-types} to every box with
$t_b>0$, obtaining
\be
-\sum_{\ell=1}^r
\sum_{\substack{b\in B_{\ell-1}}}
t_b
D_{\mathrm{KL}}
\left(
\nu_b^{(\ell)}
\middle\|
q^{<\ell}(\cdot\mid b)
\right)-C_n
\leq
\log
\mathbb P\left(
\boldsymbol N=\mathfrak n
\right)
\leq
-\sum_{\ell=1}^r
\sum_{\substack{b\in B_{\ell-1}}}
t_b
D_{\mathrm{KL}}
\left(
\nu_b^{(\ell)}
\middle\|
q^{<\ell}(\cdot\mid b)
\right).
\label{eq:array-method-of-types-bounds}
\ee

where $C_n:=\sum_{\ell=1}^r
\sum_{\substack{b\in B_{\ell-1}}}
|X_\ell|\log(t_b+1)$. We next control $C_n$. One can easily prove by induction that the total population at level
$\ell$ satisfies
\be
\sum_{b\in B_\ell}
n_b^{(\ell)}
\leq
c_\ell n. \qquad 
0\leq t_b\leq c_\ell n,
\qquad b\in B_{\ell-1}.
\label{eq:fixed-array-parent-population-bound}
\ee
Since the number of levels and boxes is fixed,
\be
\sum_{\ell=1}^r
\sum_{\substack{b\in B_{\ell-1}\\t_b>0}}
|X_\ell|\log(t_b+1)
\leq
\sum_{\ell=1}^r
|B_{\ell-1}|\,|X_\ell|
\log(c_\ell n+1)
=
O\bigl(\log(n+1)\bigr),
\label{eq:array-method-of-types-error-bound}
\ee
where the constant is independent of $\mathfrak n$. Combining \eqref{eq:array-method-of-types-error-bound} and 
\eqref{eq:array-method-of-types-bounds} one gets
\be
\begin{split}
\log
\mathbb P\left(
\boldsymbol N=\mathfrak n
\right)
&=
-\sum_{\ell=1}^r
\sum_{b\in B_{\ell-1}}
t_b
D_{\mathrm{KL}}
\left(
\nu_b^{(\ell)}
\middle\|
q^{<\ell}(\cdot\mid b)
\right)
\\
&\quad
+
O\bigl(\log(n+1)\bigr),
\end{split}
\label{eq:array-relative-entropy-expansion}
\ee
uniformly over $\mathfrak n\in\mathfrak N_n$. It remains to compare the leading term in
\eqref{eq:array-relative-entropy-expansion} with
$nI_q(p_{\mathfrak n})$. We start noticing  that since $q$ has full support,
\be
q_*
:=
\min_{\substack{1\leq\ell\leq r,\,
b\in B_{\ell-1}\\
x\in X_\ell}}
q^{<\ell}(x\mid b)
>0.
\label{eq:array-uniform-reference-lower-bound}
\ee
Now, for every probability vector $\nu$ on $X_\ell$,
\be
\begin{split}
D_{\mathrm{KL}}
\left(
\nu
\middle\|
q^{<\ell}(\cdot\mid b)
\right)
&=
\sum_{x\in X_\ell}
\nu_x\log\nu_x
-
\sum_{x\in X_\ell}
\nu_x\log q^{<\ell}(x\mid b)
\\
&\leq
-\sum_{x\in X_\ell}
\nu_x\log q^{<\ell}(x\mid b)\leq
\log\frac1{q_*}.
\end{split}
\label{eq:array-uniform-relative-entropy-bound}
\ee

By Lemma~\ref{lem:population-approximation}, there exist constants
$A_1,\ldots,A_r<\infty$, independent of $n$ and of the realizable array,
such that
\be
\left|
t_b
-
nc_\ell p_{\mathfrak n}^{(\ell-1)}(b)
\right|
\leq
A_\ell,
\qquad
b\in B_{\ell-1}.
\label{eq:array-parent-marginal-approximation}
\ee
Consequently,
\be
\begin{split}
&
\left|
\sum_{\ell=1}^r
\sum_{b\in B_{\ell-1}}
t_b
D_{\mathrm{KL}}
\left(
\nu_b^{(\ell)}
\middle\|
q^{<\ell}(\cdot\mid b)
\right)
-
nI_q(p_{\mathfrak n})
\right|
\\
&\qquad
=
\left|
\sum_{\ell=1}^r
\sum_{b\in B_{\ell-1}}
\left(
t_b
-
nc_\ell p_{\mathfrak n}^{(\ell-1)}(b)
\right)
D_{\mathrm{KL}}
\left(
\nu_b^{(\ell)}
\middle\|
q^{<\ell}(\cdot\mid b)
\right)
\right|
\\
&\qquad
\leq
\sum_{\ell=1}^r
\sum_{b\in B_{\ell-1}}
A_\ell\log\frac1{q_*}
=:C_1<\infty.
\end{split}
\label{eq:array-exponent-rate-comparison}
\ee

Combining
\eqref{eq:array-relative-entropy-expansion} and
\eqref{eq:array-exponent-rate-comparison}, we conclude that, uniformly over $\mathfrak n\in\mathfrak N_n$, 
\be
\log
\mathbb P\left(
\boldsymbol N=\mathfrak n
\right)
=
-nI_q(p_{\mathfrak n})
+
O\bigl(\log(n+1)\bigr)
\label{eq:array-probability-conclusion}
\ee

\end{proof}

We will now prove the following:

\begin{lemma}\label{lem:continuityIp}
$I(q)$ is continuous on $\mathcal{P}_X$ and is a good rate function. Moreover the unique minimum is attained at $p=q$.
\end{lemma}
\begin{proof}
For $1\leq\ell\leq r$, $b\in B_{\ell-1}$, and
$\lambda=(\lambda_x)_{x\in X_\ell}\in
\mathbb R_{\geq0}^{X_\ell}$, set
\be
\lambda_{\mathrm{tot}}
:=
\sum_{x\in X_\ell}\lambda_x
\label{eq:perspective-total-mass}
\ee
and define
\be
\Phi_\ell^b(\lambda)
:=
\begin{cases}
\lambda_{\mathrm{tot}}
D_{\mathrm{KL}}
\left(
\dfrac{\lambda}{\lambda_{\mathrm{tot}}}
\middle\|
q^{<\ell}(\cdot\mid b)
\right),
& \lambda_{\mathrm{tot}}>0,\\
0,
& \lambda_{\mathrm{tot}}=0.
\end{cases}
\label{eq:relative-entropy-perspective}
\ee
Because $q$ has full support we already observed that $q^*$ \eqref{eq:array-uniform-reference-lower-bound} is striclty greater than zero. Therefore,
\be
0
\leq
\Phi_\ell^b(\lambda)
\leq
\lambda_{\mathrm{tot}}\log\frac1{q_*}
\label{eq:perspective-bound-at-zero}
\ee
implies  that $\Phi_\ell^b$ is continuous at $\lambda=0$. Its continuity
away from zero follows from the continuity of relative entropy. Hence $\Phi_\ell^b$ is continuous on
$\mathbb R_{\geq0}^{X_\ell}$. For every $p\in\mathcal P_X$, one has $I_q(p)
=
\sum_{\ell=1}^r
c_\ell
\sum_{b\in B_{\ell-1}}
\Phi_\ell^b
\left(
p^{(\ell)}(\cdot,b)
\right).
\label{eq:rate-function-perspective-representation}$ 
Since $p\mapsto p^{(\ell)}(\cdot,b)$ is linear and continuous, this
representation shows that $I_q$ is continuous on $\mathcal P_X$. Since
$\mathcal P_X$ is compact, $I_q$ is a good rate function.

We now prove the uniqueness of the minimum of $I_q$. Every term in \eqref{eq:reinforced-rate-function} is nonnegative. If
$I_q(p)=0$, then the level-one term vanishes, and hence
\be
p^{<1}=q^{<1}.
\label{eq:zero-rate-first-level}
\ee
Since $q$ has full support, $p^{<1}$ has full support. Suppose inductively
that
\be
p^{(\ell-1)}(b)>0
\qquad
\text{for every }b\in B_{\ell-1}.
\label{eq:positive-parent-zero-rate-induction}
\ee
The vanishing of the level-$\ell$ terms then implies
\be
p^{<\ell}(\cdot\mid b)
=
q^{<\ell}(\cdot\mid b)
\qquad
\text{for every }b\in B_{\ell-1}.
\label{eq:zero-rate-conditional-kernels}
\ee
Since the kernels of $q$ have full support, it follows that
$p^{(\ell)}$ also has full support. Proceeding recursively through all
levels, all the conditional kernels of $p$ agree with those of $q$.
Therefore,
\be
p=q.
\label{eq:unique-zero-conclusion}
\ee
Conversely, $I_q(q)=0$, so $q$ is the unique zero of $I_q$.

\end{proof}

By  Lemma \ref{lem:array-probability} , given   $p\in\mathcal P_X$ and land the associated discretization 
$\mathfrak d_{n,p}\in\mathfrak N_n$ one has 

\be
\frac1n
\log
\mathbb P\left(
\boldsymbol N=\mathfrak d_{n,p}
\right)
=
-I_q(p_{\mathfrak d_{n,p}})
+
O\left(\frac{\log n}{n}\right).
\label{eq:generic-discretization-array-probability}
\ee

Therefore the continuity property of $I_q$ implies that 
\be
I_q(p_{\mathfrak d_{n,p}})
\longrightarrow
I_q(p).
\label{eq:generic-discretization-rate-convergence}
\ee
Combining
\eqref{eq:generic-discretization-array-probability} and
\eqref{eq:generic-discretization-rate-convergence}  we obtain
\be
\frac1n
\log
\mathbb P\left(
\boldsymbol N=\mathfrak d_{n,p}
\right)
=
-I_q(p)+o(1),
\label{eq:generic-point-probability-proof}
\ee
which proves \eqref{eq:generic-point-probability}.

\end{proof}

Now we start the proof of the LDP.
Let $F\subseteq\mathcal P_X$ be closed, and define
\be
\mathfrak N_n(F)
:=
\left\{
\mathfrak n\in\mathfrak N_n:
p_{\mathfrak n}\in F
\right\}.
\label{eq:realizable-arrays-in-closed-set}
\ee
Since distinct population arrays determine disjoint events,
\be
\mathbb P(\widehat p_n\in F)
=
\sum_{\mathfrak n\in\mathfrak N_n(F)}
\mathbb P\left(
\boldsymbol N=\mathfrak n
\right).
\label{eq:closed-set-array-decomposition}
\ee

The total population at level $\ell$ satisfies
\be
\sum_{b\in B_\ell}N_b^{(\ell)}
\leq
c_\ell n.
\label{eq:level-population-upper-bound}
\ee
Indeed, the assertion is immediate at level one, while for $\ell\geq2$,
\be
\sum_{b\in B_\ell}N_b^{(\ell)}
\sum_{b'\in B_{\ell-1}}T_{b'}
=
\sum_{b'\in B_{\ell-1}}
\left\lfloor
m_\ell N_{b'}^{(\ell-1)}
\right\rfloor
\leq
m_\ell
\sum_{b'\in B_{\ell-1}}
N_{b'}^{(\ell-1)}.
\label{eq:level-population-upper-bound-induction}
\ee
The result follows by induction, since
$c_\ell=m_\ell c_{\ell-1}$. In particular, every coordinate of an array at level $\ell$ belongs to
$\{0,\ldots,\lfloor c_\ell n\rfloor\}$. Consequently,
\be
|\mathfrak N_n|
\leq
\prod_{\ell=1}^r
\left(
\lfloor c_\ell n\rfloor+1
\right)^{|B_\ell|}
\leq
\prod_{\ell=1}^r
(c_\ell n+1)^{|B_\ell|}.
\label{eq:polynomial-number-realizable-arrays}
\ee
Thus the number of realizable arrays grows at most polynomially in $n$.

By the upper bound in
\eqref{eq:array-method-of-types-bounds}, there exists a constant $C_0<\infty$
such that
\be
\mathbb P\left(
\boldsymbol N=\mathfrak n
\right)
\leq
\exp\left\{
-nI_q(p_{\mathfrak n})
+
C_0\log(n+1)
\right\}
\label{eq:uniform-array-upper-bound}
\ee
for every $\mathfrak n\in\mathfrak N_n$. If
$\mathfrak n\in\mathfrak N_n(F)$, then
\be
I_q(p_{\mathfrak n})
\geq
\inf_{p\in F}I_q(p).
\label{eq:closed-set-rate-infimum}
\ee
Using \eqref{eq:closed-set-array-decomposition},
\eqref{eq:polynomial-number-realizable-arrays}, and
\eqref{eq:uniform-array-upper-bound}, we obtain
\be
\begin{split}
\mathbb P(\widehat p_n\in F)
&\leq
|\mathfrak N_n(F)|
\exp\left\{
-n\inf_{p\in F}I_q(p)
+
C_0\log(n+1)
\right\}
\\
&\leq
\prod_{\ell=1}^r
(c_\ell n+1)^{|B_\ell|}
\exp\left\{
-n\inf_{p\in F}I_q(p)
+
C_0\log(n+1)
\right\}.
\end{split}
\label{eq:closed-set-upper-bound-prelimit}
\ee
Taking logarithms, dividing by $n$, and letting $n\to\infty$, we find
\be
\limsup_{n\to\infty}
\frac1n
\log
\mathbb P(\widehat p_n\in F)
\leq
-\inf_{p\in F}I_q(p).
\label{eq:closed-set-upper-bound}
\ee

Now let $G\subseteq\mathcal P_X$ be open and fix $p\in G$. By assumption, there
exists a sequence $\mathfrak d_{n,p}\in\mathfrak N_n$ such that $p_{\mathfrak d_{n,p}} 
\longrightarrow p$ in the $\|\cdot\|_1$ norm. Since $G$ is open, it follows that $p_{\mathfrak d_{n,p}}\in G$ for all sufficiently large $n$. Thus
\be
\begin{split}
\mathbb P(\widehat p_n\in G)
&\geq
\mathbb P\left(
\boldsymbol N=\mathfrak d_{n,p}
\right).
\end{split}
\label{eq:open-set-single-array-lower-bound}
\ee
Applying  \eqref{eq:array-method-of-types-bounds}, we obtain
\be
\begin{split}
\liminf_{n\to\infty}
\frac1n
\log
\mathbb P(\widehat p_n\in G)
&\geq
\liminf_{n\to\infty}
\frac1n
\log
\mathbb P\left(
\boldsymbol N=\mathfrak d_{n,p}
\right)
\\
&=
-\lim_{n\to\infty}
I_q(p_{\mathfrak d_{n,p}})
\\
&=
-I_q(p),
\end{split}
\label{eq:open-set-lower-bound-point}
\ee
where the last equality follows from the continuity of $I_q$. Taking the
supremum over $p\in G$ gives
\be
\liminf_{n\to\infty}
\frac1n
\log
\mathbb P(\widehat p_n\in G)
\geq
-\inf_{p\in G}I_q(p).
\label{eq:open-set-lower-bound}
\ee

{
\begin{remark}
Related reinforcement mechanisms that preserve a deterministic law-of-large-numbers limit while modifying the large-deviation behavior of the empirical measure have been studied for reinforced chains in \cite{BudhirajaWaterbury2022}. A repeat-or-resample urn with the same general property also appears in the analysis of reinforced Galton--Watson processes in \cite{BertoinMallein2025III}. In both cases, the corresponding rate function is characterized variationally.

For the reinforced multinomial process considered here, the rate function takes the explicit hierarchical form \eqref{eq:reinforced-rate-function}: it is a level-dependent sum of conditional Kullback--Leibler divergences. This makes it possible to calibrate the amplification or depletion separately at each level. 
\end{remark}

\subsection{The case of random reinforcement}
\label{rem:bernoulli-thinning}
In this section we show that when the depletion of particles is random the law of large numbers can be kept unaltered, but the multiscale structure cannot be preserved.

Let $r=2$ and replace the deterministic rescaling
$T_j=\lfloor(1+\gamma)N_j^{(1)}\rfloor$ of
\S\ref{subsec:two-scale-reinforcement} by the following random reinforcement.
For each $j\in X_1$, of the $N_j^{(1)}$ only a certain fraction survives, with a certain probability. Conditionally on $N^{(1)}$, the variables $(T_j)_{j\in X_1}$ are therefore independent, with
\be
T_j\mid N^{(1)}
\sim
\operatorname{L}\bigl(\cdot\mid N_j^{(1)}\bigr)
\ee
where $\operatorname{L}$ models the depletion. Assume $\operatorname{L}$ satisfies a local
large-deviation estimate, uniformly over its realizable values, with a good rate function, denoted by $J:[0,1]\longrightarrow[0,\infty]$, that takes the survived fraction as an argument:
\begin{align}
    \operatorname{L}(T_j\mid N_j^{(1)})=\exp\big\{-N_j^{(1)}J(T_j/N_j^{(1)})+o(N_j^{(1)})\big\}.
\end{align}
Thus, at the level of conditional expectations, if we call $\theta$ the minimizer of the rate function, this random reinforcement corresponds to the deterministic rescaling with $m_2=\theta$, i.e.\
$1+\gamma=\theta$. Nevertheless, the two constructions have different rate
functions as we shall see hereby.

Let $N_{(i,j)}^{(2)}$, $i\in X_2$, denote the second-level allocation counts
among the $T_j$ surviving balls. The resulting counts satisfy
\be
\sum_{j\in X_1}N_j^{(1)}=n,
\qquad
0\leq T_j\leq N_j^{(1)},
\qquad
\sum_{i\in X_2}N_{(i,j)}^{(2)}=T_j.
\label{eq:bernoulli-count-constraints}
\ee
Conditionally on $(N_j^{(1)},T_j)_{j\in X_1}$, the allocation vectors are
independent and satisfy
\be
\bigl(N_{(i,j)}^{(2)}\bigr)_{i\in X_2}
\sim
\mathcal M\left(
T_j,q^{<2}(\cdot\mid j)
\right).
\ee

Let $(n^{(1)}_j)_{j\in X_1}$, $(t_j)_{j\in X_1}$, and
$(n^{(2)}_{i,j})_{i\in X_2,\,j\in X_1}$ realizable integer arrays that corresponds to a possible outcome of $(N^{(1)}_j)_{j\in X_1},(T_j)_{j\in X_1},(N^{(2)}_{i,j})_{(i,j)\in X}$, namely satisfying
\eqref{eq:bernoulli-count-constraints}.
By conditioning and using the conditional independence of the variables belonging to different parent boxes, we obtain
\be
\begin{split}
&
\mathbb P\left(
A_n \right):= \mathbb P\left(
N^{(1)}_j=n^{(1)}_j,\ T_j=t_j,\ N^{(2)}_{i,j}=n^{(2)}_{i,j},
\ i\in X_2,\ j\in X_1
\right)
\\
&\quad=
\mathbb P\left(N^{(1)}_j=n^{(1)}_j,\ j\in X_1\right)
\prod_{j\in X_1}
\mathbb P\left(T_j=t_j\mid N^{(1)}_j=n^{(1)}_j\right)
\\
&\qquad\quad\times
\prod_{j\in X_1}
\mathbb P\left(
N^{(2)}_{i,j}=n^{(2)}_{i,j},\ i\in X_2
\mid T_j=t_j
\right).
\end{split}
\label{eq:bernoulli-exact-factorization}
\ee

Introduce the empirical  distribution of the first level 
\be
a_j:=\frac{n^{(1)}_j}{n},
\qquad j\in X_1.
\label{eq:bernoulli-first-level-type}
\ee
Whenever $n^{(1)}_j>0$, define the empirical survival fraction by
\be
s_j:=\frac{t_j}{n^{(1)}_j}
\label{eq:bernoulli-survival-type}
\ee
and whenever $t_j>0$, define the empirical  conditional
distribution of the second level by
\be
u_j(i):=\frac{n^{(2)}_{i,j}}{t_j},
\qquad i\in X_2.
\label{eq:bernoulli-second-level-type}
\ee
For $t_j>0$, set also 
\begin{align}
d_j:=
D_{\mathrm{KL}}
\left(
u_j
\middle\|
q^{<2}(\cdot\mid j)
\right).
\end{align}
By standard large deviation estimates on multinomial, the first-level allocation
satisfies
\be
P\left(N^{(1)}_j=n^{(1)}_j,\ j\in X_1\right)
=
\exp\left\{
-nD_{\mathrm{KL}}
\left(
a\middle\|q^{<1}
\right)
+O(\log n)
\right\}.
\label{eq:bernoulli-first-level-cost}
\ee

The $\operatorname{L}$--LDP instead yields
\be
\mathbb P\left(T_j=t_j\mid N^{(1)}_j=n^{(1)}_j\right)
=
\exp\left\{
-n a_jJ(s_j)
+o\bigl(n\bigr)
\right\}.
\label{eq:bernoulli-survival-cost}
\ee
To fix ideas, let us keep in mind the example of a binomial depletion process, where each ball after the first allocation survives with probability $\theta$. This clearly satisfies the requirements on $\operatorname{L}$ by the strong law of large numbers, and the rate function would be the binomial cross entropy
\be
J_{\rm bin.}(s)
:=
s\log\frac{s}{\theta}
+
(1-s)\log\frac{1-s}{1-\theta},
\qquad s\in[0,1].
\label{eq:bernoulli-relative-entropy}
\ee

Conditionally on $T_j=t_j$, the second-level allocation follows a multinomial distribution and then, once more:
\be
\mathbb P\left(
N^{(2)}_{i,j}=n^{(2)}_{i,j},\ i\in X_2
\mid T_j=t_j
\right)=
\exp\left\{
-t_jD_{\mathrm{KL}}
\left(
u_j
\middle\|
q^{<2}(\cdot\mid j)
\right)
+O\bigl(\log(n)\bigr)
\right\}.
\label{eq:bernoulli-second-level-cost}
\ee
Using $t_j=s_j n^{(1)}_j=na_js_j$ and the definition of $d_j$ one obtains an estimate for \eqref{eq:bernoulli-exact-factorization}
\be
\begin{split}
\mathbb P(A_n)
=
\exp\Bigg\{
-nD_{\mathrm{KL}}
\left(
a\middle\|q^{<1}
\right)
-n\sum_{j\in X_1}
a_j
\left[
J(s_j)+s_jd_j
\right]
+o(n)
\Bigg\}.
\end{split}
\label{eq:bernoulli-joint-type-cost}
\ee

Take $p \in \mathcal{P}_X$ such that $p^{<1}(j)=a_j$ 
and $p^{<2}(i|j)=u_j(i)$.
The survival fraction $s_j$ is determined by the complete
second-level count array, since
$T_j=\sum_{i\in X_2}N_{i,j}^{(2)}$, but it is not determined by the probability vector $p$  which retains only the normalized
conditional kernel. Therefore, contracting the joint rate function over the unobserved survival fractions gives a rate function 
\be
I(p)
=
D_{\mathrm{KL}}
\left(
p^{<1}\middle\|q^{<1}
\right)
+
\sum_{j\in X_1}
a_j
\Psi_J(d_j),
\label{eq:bernoulli-contraction}
\ee
where
\be
\Psi_J(d):=\inf_{s\in[0,1]}\{J(s)+sd\}.
\label{eq:generic-effective-cost}
\ee
Since $J$ has the unique zero $\theta>0$, one has
$\Psi_J(0)=0$ and $\Psi_J(d)>0$ for every $d>0$. Consequently,
$I(p)=0$ if and only if $p=q$. Thus the random depletion preserves
the same law-of-large-numbers limit as the deterministic rescaling,
while changing its large-deviation cost.
Let us evaluate this cost in our Bernoulli running example.

For the Bernoulli depletion process, a direct computation gives 
\be
\inf_{s\in[0,1]}
\left\{
J_{\rm bin.}(s)+sd
\right\}
=
\Psi_\theta(d),
\qquad
\Psi_\theta(d)
:=
-\log\left(1-\theta+\theta e^{-d}\right).
\label{eq:bernoulli-effective-cost}
\ee
Inserting this identity into \eqref{eq:bernoulli-contraction} one gets the rate function of the process with random reinforcement
\be\label{eq:rate-bernoulli}
I(p)=D_{\rm KL}\big(p^{<1}\|q^{<1}\big)
+\sum_{j\in X_1}p^{<1}(j)\,\Psi_\theta\big(d_j\big).
\ee
Since $\Psi_\theta$ is not linear, \eqref{eq:rate-bernoulli} is not of the imbalanced
form required by Proposition~\ref{prop:var_princ_Maxent}. The deviation from the
deterministic rate function $I_{\rm det}$ with the same mean $m_2=\theta$ is exact and
explicit,
\be\label{eq:bernoulli-defect}
I(p)-I_{\rm det}(p)=\sum_{j\in X_1}p^{<1}(j)\Big[\Psi_\theta(d_j)-\theta\,d_j\Big]\;\leq\;0,
\ee
with equality only when $p^{<2}(\cdot\mid j)=q^{<2}(\cdot\mid j)$ at every $j$ of
positive mass. Writing $p=q+\varepsilon \delta $ with $\sum_x \delta(x)=0$, and setting
$\chi^2(x_1)=\sum_{x_2}v^{<2}(x_2\mid x_1)^2/q^{<2}(x_2\mid x_1)$ with
$v^{<2}=\partial_\varepsilon|_0 p^{<2}$, one has $d_j=\tfrac12\varepsilon^2\chi^2(j)+O(\varepsilon^3)$
and $\Psi_\theta(d)-\theta d=-\tfrac12\theta(1-\theta)d^2+O(d^3)$, whence
\be\label{eq:bernoulli-fourth}
I(p)-I_{\rm det}(p)=-\frac{\theta(1-\theta)}{8}\,\varepsilon^4
\sum_{x_1}q^{<1}(x_1)\,\chi^2(x_1)^2+O(\varepsilon^5).
\ee

The defect is therefore of \emph{fourth} order. Indeed, the randomness of the
surviving population changes the effective cost $\Psi_\theta(d)$ only at
second order in $d$, while $d$ is a relative entropy and is itself quadratic
in the deviation from $q$.

\section{Multiscale measures and Poisson--Dirichlet processes}
\label{sec:chinese}

Poisson--Dirichlet processes are classical objects that appear also in the description of the asymptotic Gibbs measure of mean-field spin glasses \cite{MPV,Guerra2002BrokenRS,Tala_vol1,panchenko2015sherrington}. In that setting, they provide the building blocks of the Derrida--Ruelle cascades
\cite{REM_Derrida,Ruelle}, which encode the random weights of the clusters in
the ultrametric decomposition of the Gibbs measure. More generally, Ruelle
probability cascades provide a probabilistic representation of multiscale
recursions \cite{panchenko2015sherrington,MultiscaleSK_CM2018}. We recall this
representation in the two-scale case and then relate it to a Chinese restaurant
process (CRP) followed by an ordinary multinomial sampling. The Poisson--Dirichlet
representation and the Chinese restaurant construction used below are
classical. Here instead we show that the multiscale measure can emerge as a combination of the CRP process with a multinomial sampling. This mechanism differs from that of Section \ref{sec:large-deviation-perspective}, since the reinforcement is given by the preferential attachment, characteristic of the CRP, on clusters rather than samples.

\subsection{Poisson--Dirichlet representation of a two-scale measure}

Consider a two-scale measure on $X=X_2\times X_1$ with parameters
$\zeta_2=1$ and $\zeta_1=\zeta\in(0,1)$. In this section it is convenient to
equip $X_1$ and $X_2$ with a priori probability measures and to denote the
corresponding averages by $\E_1$ and $\E_2$. For uniform a priori measures,
this convention differs from the use of counting measures in
Definition~\ref{def:multiscale measure} only by additive constants in the
pressures and leaves the associated probability measure unchanged. Set
\be
Z(x_1):=\E_2\,e^{H(x_2,x_1)}.
\label{eq:section-five-random-partition}
\ee
The two-scale pressure is then
\be
P_0=\frac{1}{\zeta}\log\E_1 Z^{\zeta}.
\label{eq:section-five-two-scale-pressure}
\ee

Let $(\nu_\alpha)_{\alpha\geq1}$ have the Poisson--Dirichlet distribution
$PD(\zeta)$, and let $(y_\alpha)_{\alpha\geq1}$ be i.i.d.\ copies of $x_1$,
independent of $(\nu_\alpha)_{\alpha\geq1}$. Writing
$Z_\alpha:=Z(y_\alpha)$, the classical Poisson--Dirichlet invariance identity
gives
\be
P_0
=\mathbb E\log\sum_{\alpha\geq1}\nu_\alpha Z_\alpha.
\label{RPC}
\ee
Here and below, $\mathbb E$ denotes expectation with respect to both the
Poisson--Dirichlet weights and the i.i.d.\ samples. Identity~\eqref{RPC} is a
standard ingredient in the theory of Ruelle probability cascades; see, for
instance, \cite{Tala_vol1,panchenko2015sherrington,MultiscaleSK_CM2018}.

Equation~\eqref{RPC} also has a simple statistical-mechanical interpretation.
Indeed, writing $h_\alpha=\log\nu_\alpha$, its right-hand side is the
quenched pressure of a system indexed by $\alpha\in\N$, with random activities
$e^{h_\alpha}=\nu_\alpha$ and partition functions $Z_\alpha$. Thus the
Poisson--Dirichlet weights play the role of a random chemical potential for the
coarse level of the two-scale system.

It is also useful to recall the behavior of this representation at the two
extreme values of $\zeta$ \cite{Tala_vol1}. When $\zeta$ is close to $1$, the
weights are diffuse and the sum $\sum_\alpha\nu_\alpha Z_\alpha$ is close, in
the appropriate sense, to $\E_1 Z$. When $\zeta$ is close to $0$, most of the
mass is carried by the largest weight, and the right-hand side of
\eqref{RPC} approaches $\mathbb E\log Z$. These are precisely the two limits
interpolated by the power mean in
\eqref{eq:section-five-two-scale-pressure}.

The same representation and invariance hold at the level of observables. For a realization
of $(\nu_\alpha,y_\alpha)_{\alpha\geq1}$, define a random probability measure
on $X_2\times\N$ by
\be
\mu(x_2,\alpha)
:=
\frac{\nu_\alpha Z_\alpha}
{\sum_{\beta\geq1}\nu_\beta Z_\beta}
\frac{e^{H(x_2,y_\alpha)}}{Z_\alpha},
\quad
\langle f\rangle^*
:=\sum_{\alpha\geq1}\E_2\bigl[
\mu(x_2,\alpha)f(x_2,y_\alpha)\bigr].
\label{eq:twoscalesRPC}
\ee
Here $\mu(x_2,\alpha)$ is understood as a density with respect to the a priori
measure on $X_2$ and counting measure on $\N$. On the other hand, the average
of $f$ with respect to the original two-scale measure is
\be
\langle f\rangle
=\E_1\left[
\frac{Z^\zeta}{\E_1 Z^\zeta}
\E_2\left(\frac{e^H}{Z}f\right)
\right].
\label{eq:twoscales2}
\ee
Applying \eqref{RPC} to the perturbed Hamiltonian $H+tf$ and differentiating
at $t=0$ yields
\be
\begin{split}
\langle f\rangle
&=\mathbb E\langle f\rangle^*
\\
&=\mathbb E\sum_{\alpha\geq1}
\frac{\nu_\alpha Z_\alpha}
{\sum_{\beta\geq1}\nu_\beta Z_\beta}
\E_2\left(
\frac{e^{H(x_2,y_\alpha)}}{Z_\alpha}
f(x_2,y_\alpha)
\right).
\end{split}
\label{eq:twoscales3}
\ee
Thus the two-scale measure is the expectation of the random Gibbs measure
$\mu$ defined in \eqref{eq:twoscalesRPC}.

\subsection{Chinese restaurant representation}

The Poisson--Dirichlet weights in \eqref{RPC} can be obtained from a Chinese
restaurant process (CRP). We recall the construction in the particular case needed
here \cite{Pitman-Yor,libro_PC}. The first customer occupies the first table.
Suppose that, after $n$ customers, there are $K_n$ occupied tables with
occupancies $n_1,\ldots,n_{K_n}$. Customer $n+1$ starts a new table with
probability $\zeta K_n/n$, while it joins an occupied table $\alpha$ with
probability
\be
\mathbb P\bigl(n+1\text{ joins table }\alpha\mid n_1,\ldots,n_{K_n}\bigr)
=\frac{n_\alpha-\zeta}{n}.
\label{eq:section-five-crp-rule}
\ee
The table labels are assigned in order of appearance. This is a classical
preferential-reinforcement mechanism: the probability of returning to an
occupied table increases with its current occupancy.

For every fixed $\alpha$, the empirical table frequencies satisfy
\be
\rho_\alpha
:=\lim_{n\to\infty}\frac{n_\alpha}{n}
\qquad\text{a.s.}
\label{RPCconv}
\ee
The sequence $(\rho_\alpha)_{\alpha\geq1}$, in order of appearance, has the
$GEM(\zeta)$ (Griffiths--Engen--McCloskey) distribution. Its decreasing rearrangement has the
$PD(\zeta)$ distribution \cite{Pitman-Yor}. Since the samples $y_\alpha$ are
i.i.d.\ and independent of the weights, the random sums and measures above
may equivalently be written using the $GEM(\zeta)$ ordering,
provided that weights and samples are relabelled together.

This gives a finite approximation of the random measure
\eqref{eq:twoscalesRPC}. Run the Chinese restaurant process for $n$ customers,
set
\be
\rho_\alpha(n):=\frac{n_\alpha}{n},
\qquad 1\leq\alpha\leq K_n,
\label{eq:section-five-empirical-crp-weights}
\ee
and attach to each occupied table an independent mark $y_\alpha$ with the law
of $x_1$. Then define
\be
\mu^{(n)}(x_2,\alpha)
:=
\frac{\rho_\alpha(n)Z_\alpha}
{\sum_{\beta=1}^{K_n}\rho_\beta(n)Z_\beta}
\frac{e^{H(x_2,y_\alpha)}}{Z_\alpha},
\qquad 1\leq\alpha\leq K_n.
\label{eq:section-five-finite-random-measure}
\ee
Again, this is a density with respect to the a priori measure on $X_2$.
The almost-sure convergence of the CRP frequencies in \eqref{RPCconv}, in
$\ell^1$, implies the convergence of $\mu^{(n)}$ to the corresponding random
measure with $GEM(\zeta)$ weights. After a simultaneous relabelling of weights
and marks, the limit has the same law as the measure $\mu$ in
\eqref{eq:twoscalesRPC}.

We can now express the same construction in the large-deviation language used
in Section~\ref{sec:large-deviation-perspective}. For simplicity, take the a
priori measure on $X_2$ to be uniform. Conditionally on the CRP frequencies,
draw an independent sample of $n$ pairs in
$X_2\times\{1,\ldots,K_n\}$ with probabilities
\be
q_n(x_2,\alpha)=q^{<1}_n(\alpha)q^{<2}_n(x_2\mid \alpha)
:=\rho_\alpha(n)\cdot\frac{1}{|X_2|}.
\label{eq:section-five-random-base-measure}
\ee
Let $Y=(Y_{x_2,\alpha})$ be the resulting multinomial count vector, and let
$p$ be a realizable empirical distribution supported on the occupied tables.
The method of types gives, conditionally on $\rho(n)$,
By the chain rule for relative entropy,
\be\label{eq:cspr}
\begin{aligned}
\frac1n\log\mathbb P\bigl(Y=np\mid\rho(n)\bigr)
&=-D_{\mathrm{KL}}(p\|q_n)+o(1)
\\
&=-D_{\mathrm{KL}}\bigl(p^{<1}\|\rho(n)\bigr)
-\sum_{\alpha=1}^{K_n}p^{<1}_\alpha
D_{\mathrm{KL}}\left(
p^{<2}_{\cdot\mid\alpha}
\middle\|
\operatorname{Unif}(X_2)
\right)
+o(1)
\\
&=-D_{\mathrm{KL}}\bigl(p^{<1}\|\rho(n)\bigr)
+\sum_{\alpha=1}^{K_n}p^{<1}_\alpha
\mathcal S\left[p^{<2}_{\cdot\mid\alpha}\right]
-\log|X_2|+o(1)
\\
&=\mathcal S[p]
+\sum_{\alpha=1}^{K_n}p^{<1}_\alpha
\log\rho_\alpha(n)
-\log|X_2|+o(1).
\end{aligned}
\ee
Although the number of occupied tables is random, the remainder is still
$o(1)$ almost surely: the standard CRP asymptotics give
$K_n/n^\zeta\to S_\zeta<\infty$ almost surely \cite{Pitman-Yor}, while the method-of-types error divided by $n$ is of order $K_n\log(n+1)/n$.

Tilting this multinomial experiment by the energy $H$ leads to the variational
problem
\be
\sup_p\left\{
\mathcal S[p]
+\sum_{x_2,\alpha}p(x_2,\alpha)H(x_2,y_\alpha)
+\sum_{\alpha=1}^{K_n}p^{<1}_\alpha\log\rho_\alpha(n)
\right\}.
\label{eq:section-five-random-variational-principle}
\ee
This is the ordinary Gibbs variational principle with the random chemical
potential $\log\rho_\alpha(n)$. Its unique maximizer is
\be
\begin{split}
p_n^*(x_2,\alpha)
&=\frac{\rho_\alpha(n)|X_2|^{-1}e^{H(x_2,y_\alpha)}}
{\sum_{\beta=1}^{K_n}\rho_\beta(n)Z_\beta}
\\
&=\frac{\rho_\alpha(n)Z_\alpha}
{\sum_{\beta=1}^{K_n}\rho_\beta(n)Z_\beta}
\frac{|X_2|^{-1}e^{H(x_2,y_\alpha)}}{Z_\alpha}.
\end{split}
\label{eq:section-five-finite-maximizer}
\ee
This is precisely the probability measure whose density with respect to the
uniform measure on $X_2$ is given by
\eqref{eq:section-five-finite-random-measure}. In the limit $n\to\infty$, it
therefore converges, up to a simultaneous relabelling, to the random measure
appearing in \eqref{eq:twoscalesRPC}.

\begin{remark}
The preceding construction is related to, but distinct from, the mechanism of
Section~\ref{sec:large-deviation-perspective}. The Poisson--Dirichlet identity,
the Chinese restaurant process, and the convergence of its empirical
frequencies to $GEM(\zeta)$ weights are classical. The observation relevant
here is that they locate reinforcement at a different part of the
construction. In Section~\ref{sec:large-deviation-perspective}, reinforcement
changes the number of samples available at successive levels of a fixed finite
hierarchy. It leaves the law-of-large-numbers limit unchanged, but modifies the
coefficients of the conditional relative entropies in the rate function. In
the Chinese restaurant process, reinforcement instead acts on the coarse
cluster occupancies and produces genuinely random limiting weights. Given
those weights, the sampling inside the clusters is an ordinary multinomial
experiment, as shown in \eqref{eq:cspr}. Thus the scale parameter $\zeta$ is
encoded here in the law of the random chemical potential
$\log\rho_\alpha$, rather than in a deterministic multiplier of a conditional
entropy. The two constructions therefore lead to the same two-scale structure
through different probabilistic mechanisms.
\end{remark}

}

\section{Conclusions}

In this work we have revisited multiscale Gibbs measures from complementary variational and probabilistic perspectives. The key observation is that their characteristic imbalance among conditional entropies can be understood as an uneven allocation of statistical information across the hierarchy. We have shown that these measures are the unique maximizers of a generalized maximum-entropy principle with constraints on conditional entropies (Proposition~\ref{prop:var_princ_Maxent}). In the two-level case, this variational formulation also recovers the thermodynamic interpretation in terms of two effective temperatures and the corresponding level-dependent response relations \cite{Kurchan-Cugliandolo,leuzzi2007thermodynamics,Vulpiani17}.

Our main contribution is the hierarchical reinforced multinomial process introduced in Section~\ref{sec:large-deviation-perspective}. By amplifying or depleting the effective population between successive levels, the process leaves the law-of-large-numbers limit unchanged while modifying the exponential cost of fluctuations. Its empirical measure satisfies a large-deviation principle with an explicit rate function given by a weighted sum of conditional Kullback--Leibler divergences (Theorem~\ref{thm:reinforced-multinomial-ldp}); for a uniform base measure, the corresponding energetic tilt reproduces the multiscale variational principle. 
The present construction provides an explicit rate function and extends this mechanism to an arbitrary hierarchy of scales. We have also shown that the deterministic character of the population rescaling is important: random rescaling introduces an additional large-deviation cost and does not, in general, produce the required multiscale structure.

Finally, we compared this mechanism with the classical Poisson--Dirichlet representation of two-scale measures and its realization through the Chinese restaurant process \cite{Pitman-Yor,libro_PC}. In that construction, familiar from the Ruelle-cascade description of mean-field spin glasses \cite{MPV,Guerra2002BrokenRS,Tala_vol1,panchenko2015sherrington,MultiscaleSK_CM2018}, reinforcement acts on cluster occupancies and produces random limiting weights. In the reinforced multinomial process, it instead changes the sampling depth along a fixed hierarchy while preserving a deterministic typical measure. The two constructions thus provide distinct probabilistic routes to the same two-scale structure: through a random chemical potential in one case, and through deterministic reweighting of conditional entropies in the other.


\paragraph*{Acknowledgements}
The authors thank Jorge Kurchan for insightful discussions. PC and EM were supported by the EU H2020 ICT48 project Humane AI Net contract number 952026; by the Italian Extended Partnership PE01 - FAIR Future Artificial Intelligence Research - Proposal code PE00000013 under the MUR National Recovery and Resilience Plan; by the project PRIN 2022 - Proposal code: J53D23003690006. The work is partially supported by GNFM (Indam), EU H2020 ICT48 project Humane AI Net.

\section*{Declarations}
The authors declare no conflict of interest. No data are associated with this manuscript.

\bibliography{bibliography}

\end{document}